\documentclass{elsart}
\usepackage{graphicx}

\usepackage{amssymb}

\newcommand{\lsim}{\mathrel{\mathop{\kern 0pt \rlap
  {\raise.2ex\hbox{$<$}}}
  \lower.9ex\hbox{\kern-.190em $\sim$}}}
\newcommand{\gsim}{\mathrel{\mathop{\kern 0pt \rlap
  {\raise.2ex\hbox{$>$}}}
  \lower.9ex\hbox{\kern-.190em $\sim$}}}

\begin{document}
\begin{flushright}
\large
{\bf ROM2F/2008/03 \\}
{\bf March 2008 -- r1\\}
\end{flushright}

\normalsize

\begin{frontmatter}

\title{The DAMA/LIBRA apparatus}

\author[d2,i2]{R. Bernabei}
\author[i2]{P. Belli}
\author[i2]{A.\,Bussolotti}
\author[d1,i1]{F.\,Cappella}
\author[i3]{R.\,Cerulli}
\author[c1]{C.J.\,Dai}
\author[d1,i1]{A. d'Angelo}
\author[c1]{H.L.\,He}
\author[i1]{A.\,Incicchitti}
\author[c1]{H.H.\,Kuang}
\author[c1]{J.M.\,Ma}
\author[i1]{A.\,Mattei}
\author[d2,i2]{F.\,Montecchia}
\author[d2,i2]{F.\,Nozzoli}
\author[d1,i1]{D.\,Prosperi}
\author[c1]{X.D.\,Sheng}
\author[c1,c2]{Z.P.\,Ye}

\address[d2]{Dip. di Fisica, Universit\`a di Roma ``Tor Vergata'', I-00133 Rome, Italy}
\address[i2]{INFN, sez. Roma ``Tor Vergata'', I-00133 Rome, Italy}
\address[d1]{Dip. di Fisica, Universit\`a di Roma ``La Sapienza'', I-00185 Rome, Italy}
\address[i1]{INFN, sez. Roma, I-00185 Rome, Italy}
\address[i3]{Laboratori Nazionali del Gran Sasso, I.N.F.N., Assergi, Italy}
\address[c1]{IHEP, Chinese Academy, P.O. Box 918/3, Beijing 100039, China}
\address[c2]{University of Jing Gangshan, Jiangxi, China}

\begin{abstract}
The $\simeq$ 250 kg highly radiopure NaI(Tl) DAMA/LIBRA 
apparatus, running at the Gran Sasso National Laboratory (LNGS) of the I.N.F.N., is described.
\end{abstract}

\end{frontmatter}

\section{Introduction}
\vspace{-0.4cm}

While running the first generation DAMA/NaI apparatus ($\simeq$ 100 kg highly radiopure NaI(Tl)) 
\cite{pro,Nim98,Sist,RNC,ijmd,ijma,epj06,ijma07,chan,wimpele,LDM,rare},
DAMA proposed to realize a ton apparatus in 1996 \cite{IDM96}. Thus, 
a second generation R\&D project for highly radiopure NaI(Tl) 
detectors was funded and carried out over several years with leader companies in order to realize, 
as a new step, a second generation experiment with an exposed mass of about 250 kg.
The $\simeq$ 250 kg highly radiopure NaI(Tl) DAMA/LIBRA (Large sodium 
Iodide Bulk for RAre processes) apparatus has been designed and built mainly to further 
investigate 
the Dark Matter 
particle component(s) in the galactic halo by exploiting the annual modulation signature;
it also aims to improve the investigation of several rare processes.
\begin{figure}[!htbp]
\centering
\includegraphics[width=12.9cm] {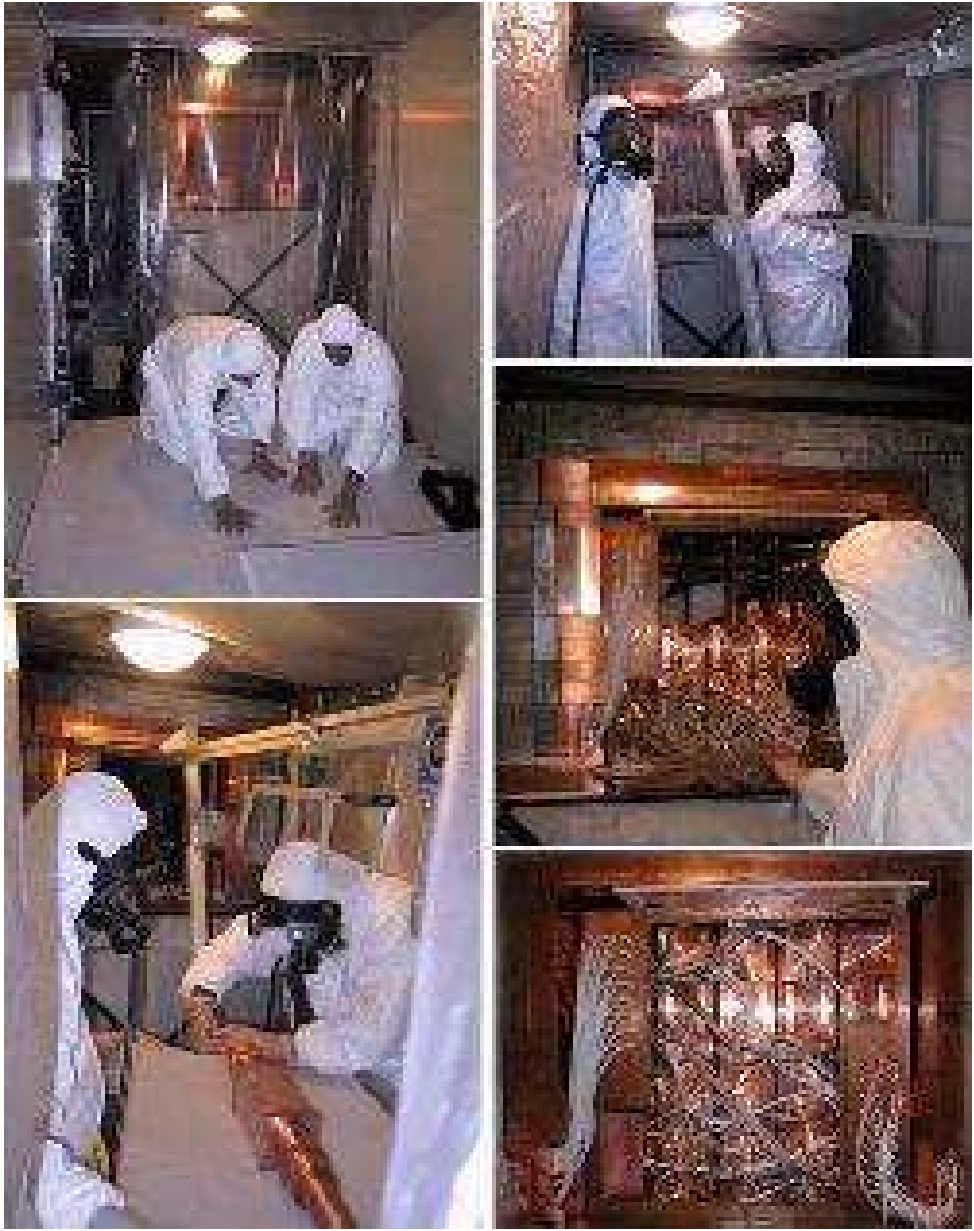}
\caption{Disinstalling DAMA/NaI and installing DAMA/LIBRA; all procedures involving 
detectors, photomultipliers (PMTs), etc.~have been carried out in HP (High Purity) Nitrogen atmosphere.
Improving the experimental site (walls, floor, etc) after removal 
of the DAMA/NaI electronics ({\it top left}); 
dismounting the DAMA/NaI detectors in HP Nitrogen atmosphere ({\it top right});
assembling a DAMA/LIBRA detector in HP Nitrogen atmosphere ({\it bottom left}); 
installing the DAMA/LIBRA detectors in HP Nitrogen atmosphere ({\it center right}) and a view of
the inner Cu box filled by the DAMA/LIBRA detectors in HP Nitrogen atmosphere ({\it bottom right}).}
\label{assem}
\end{figure}
DAMA/NaI completed its data taking in July 2002  
and the installation of DAMA/LIBRA started (see Fig.~\ref{assem}).
DAMA/LIBRA has begun first operations in March 2003. 

In the present paper the DAMA/LIBRA apparatus
and, in particular, the features of its new detectors, developed 
by means of the second generation R\&D, are described.

\vspace{-0.6cm}
\section{The target material}
\vspace{-0.8cm}

Highly radiopure NaI(Tl) scintillators have been chosen as target -- detector material for particle
Dark Matter investigations since they offer many competitive aspects; among them we remind:
 1) well known technology;
 2) high radiopurity reachable by material selections and protocols, by chemical/physical 
    purifications, etc.;
 3) large mass feasible;
 4) high duty cycle;
 5) routine calibrations feasible down to keV range in the same conditions as the
    production runs;
 6) well controlled operational conditions and monitoring feasible;
 7) absence of microphonic noise;
 8) suitable signal/noise discrimination near the energy threshold 
    profiting of the relatively high available number 
    of photoelectrons/keV and of the well different timing structures of the PMT noise pulses 
    (single fast photoelectrons with decay time of order of tens of ns) with respect to the 
    NaI(Tl) scintillation pulses (time distribution with time decay of order of hundreds of ns);
 9) high light response, that is keV threshold really reachable;
10) no necessity of re-purification or cooling down/warming up procedures (implying high 
    reproducibility, high stability, etc.);
11) possibility to exploit the granularity of the apparatus, an interesting feature for Dark 
Matter particle
    investigations and for background recognition;
12) sensitivity to both high (by Iodine target) and low (by Na target) mass WIMP (Weakly Interacting 
Massive Particles) candidates;
13) high sensitivity to the class of WIMP candidates with spin-independent (SI), spin-dependent 
    (SD) and mixed (SI\&SD) couplings (see e.g. \cite{RNC,bb});
14) high sensitivity to several other existing scenarios (see e.g. 
    \cite{RNC,ijmd,ijma,epj06,ijma07,chan,wimpele,LDM,oo}) and
    to many other possible candidates including those producing just 
    electromagnetic radiation in the interaction;
15) possibility to effectively investigate the annual modulation signature in all the
    needed aspects; 
16) ``ecologically clean'' apparatus, thus no safety problems;
17) technique cheaper than every other considered in the field; 
18) small underground space needed;
19) pulse shape discrimination feasible at reasonable level when of interest.

Moreover, highly radiopure NaI(Tl) scintillators can also offer the possibility to achieve significant 
results on several other rare processes -- as already done by the DAMA/NaI apparatus -- such as e.g. 
\cite{Okun}:
i)    possible violation of Pauli exclusion principle in $^{127}$I and in $^{23}$Na;
ii)   electron stability;
iii)  charge non-conserving processes;
iv)   search for solar axions;
v)    search for exotic matter;
vi)   search for spontaneous transition of nuclei to a superdense state;
vii)  search for spontaneous emission of heavy clusters in $^{127}$I;
viii) search for possible nucleon, di-nucleon and tri-nucleon decays into invisible channels; etc.

\section{Layout of the experimental apparatus}

The main parts of the experimental apparatus are:
i)  the installation;
ii) the multicomponent passive shield;
iii) the 25 highly radiopure NaI(Tl) detectors;
iv) the glove-box for calibrations;
v) the electronic chain and the monitoring/alarm system; 
vi) the data acquisition (DAQ) system.

All the materials constituting the apparatus have been selected for low radioactivity by measuring 
samples 
with low background Ge detector deep underground in the Gran Sasso National Laboratory and/or 
by Mass Spectrometry (MS) and by Atomic Absorption Spectroscopy (AAS).

\subsection{The installation}

\begin{figure}[!htbp]
\centering
\includegraphics[width=11.cm] {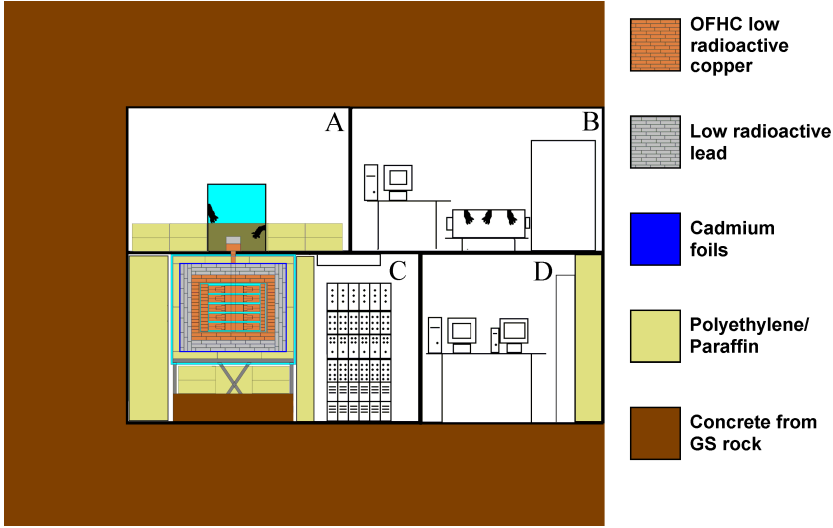}
\caption{Schematic view of the building hosting the DAMA/LIBRA apparatus  (not to scale).
About 1 m concrete (made from the Gran Sasso rock material) almost fully surrounds 
(mostly outside the building) the apparatus. In particular, the room $A$ hosts the glove box for 
calibrations;
some facilities are in the room $B$; 
the sensitive parts of the apparatus are in the room $C$ inside the multi-component shield as well as
the electronics; the computer for DAQ is in the room $D$.}
\label{setup}
\end{figure}

A schematic view of the building housing the DAMA/LIBRA experimental apparatus is shown in 
Fig.~\ref{setup}.
Even the materials, the paints, etc.,  used to build the installation, have been selected 
for low radioactivity as much as possible at time of construction. Most of the materials have been underground since 
about 15 years or more.

At the ground floor, the room $D$ in Fig.~\ref{setup} houses the computer devoted to the data acquisition and 
the room $C$ contains the sensitive part of the apparatus, the passive shield, the electronic 
chain and the acquisition system. The floor and the walls of the room $C$ 
are sealed from external air by a tight coverage made by Supronyl 
(permeability equal to $2 \times 10^{-11}$ cm$^2$/s \cite{woj91});
this is the first level of sealing from the Radon present in trace in the air of the underground laboratory.
The door at entrance of the room $C$ is air tight and -- since HP Nitrogen is released in this room --
a low level oxygen alarm is in operation.
The level of the environmental Radon inside the experimental room $C$ is continuously monitored by a Radon-meter and 
recorded with the production data; it shows that the environmental Radon in the installation
is at level of the Radon-meter sensitivity ($\simeq 3 $ Bq/m$^3$).

The DAMA/LIBRA passive shield is supported by a metallic structure mounted above a concrete basement (see Fig.~\ref{setup}).
A neoprene layer separates the concrete basement and the floor of the laboratory.
The space between this basement and the metallic structure is filled by paraffin for several tens cm in height.
This shield is sealed in a plexiglass box maintained in HP Nitrogen atmosphere
in slight overpressure with respect to its external environment; this is the second level of 
sealing from the external air.

Two mechanical systems allow the lowering of the front and the back sides of the shield 
through an electric engine
and, hence, to access the detectors during their installation (also see Fig. \ref{assem}).

On the top of the shield (at the first floor, room $A$) a glove-box (also continuously maintained in the HP Nitrogen 
atmosphere) is directly connected through
Cu pipes to the inner Cu box, housing the detectors. 
The inner Cu box is also continuously flushed with HP Nitrogen and maintained in slight overpressure 
with respect to its external environment (this is the third level of sealing from external 
air in the underground laboratory). 
The pipes are filled with low radioactivity Cu bars (covered by 10 cm of low 
radioactive Cu and 15 cm of low radioactive Pb) which can be removed to allow the insertion 
of radioactive sources holders for calibrating the detectors in the same running condition without 
any contact with the installation environment. The glove-box is also equipped with a compensation
chamber.

Paraffin/polyethylene fill the space surrounding the plexiglass box as much as possible depending on
the available space. Moreover, 
mostly outside the building, the DAMA/LIBRA apparatus is almost fully surrounded by about 1 m 
concrete 
made from the Gran Sasso rock material.

The whole installation is under air conditioning to assure a suitable and stable working temperature for 
the electronics. The huge heat capacity of the multi-tons passive shield ($\approx 10^6$ cal$/ ^{\circ}$C) 
assures further a relevant stability of the detectors' operating temperature.
In particular, two independent systems of air conditioning are available for redundancy: 
one cooled by water refrigerated by a devoted chiller and the other operating with cooling gas.

\begin{figure}[!t]
\centering
\includegraphics[width=12.cm] {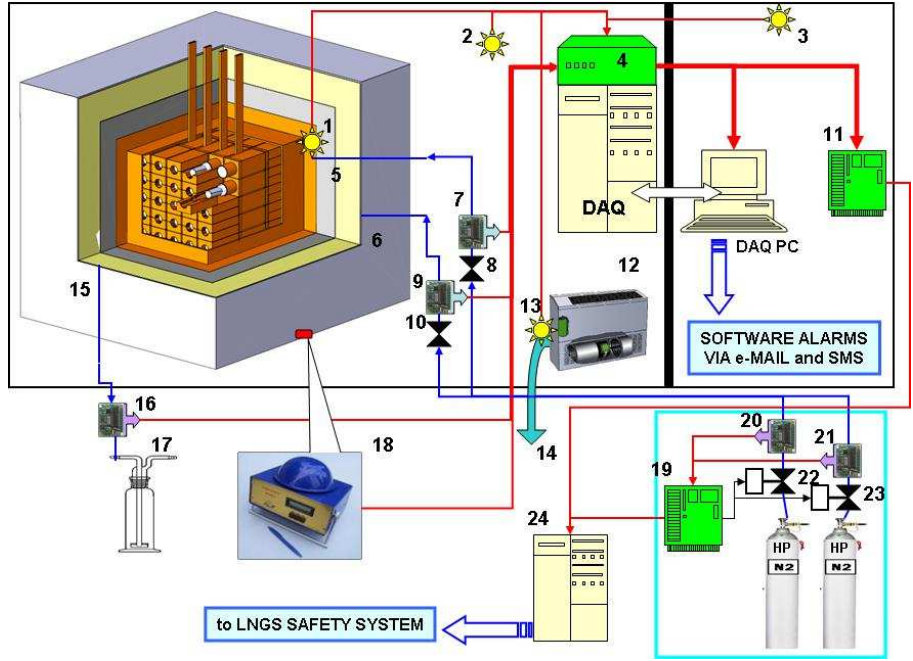}
\caption{A schematic and simplified view of the monitoring/alarm system.
{\em Legend:} 
1) temperature probe of the detectors' environment;
2) temperature probes of the electronic devices;
3) temperature probe of the DAQ system;  
4) interface of the probes;
5) HP N$_2$ gas inlet in the inner CU box;
6) HP N$_2$ gas inlet in the external plexiglass box;
7) and 9) fluxmeters;
8) and 10) valves for the flux regulating;
11) interface towards the LNGS alarm system;
12) fan coil;
13) temperature probe for the water in the cooling system;
14) cooling system towards a chiller;
15) gas outlet;
16) pressure gauge;
17) gas bubbler;
18) Radon meter;
19) interface of the control of the gas bottles;
20) and 21) pressure meters;
22) and 23) electrovalves;
24) local junction box of the safety and alarm system of LNGS.
Moreover, self-controlled computer processes automatically monitor also several other parameters (the above mentioned and
those from the DAQ) and manage software alarms (by e-mails and by SMS) 
to operator.}
\label{fg:proc}
\end{figure}

A hardware/software monitoring system provides data on the operating conditions. 
In particular, several probes are read out and the results are stored 
with the production data. A schematic view of the monitoring/alarm system is reported in 
Fig.~\ref{fg:proc}. It is directly interfaced to the safety system of the laboratory, for prompt action
of the LNGS operator. Moreover, self-controlled computer processes automatically monitor several parameters 
-- including those from DAQ -- and manage alarms (by e-mails and by SMS) to the DAMA operator 
if one of them would go out of the allowed range
(whose restrictions depend on the required stability level of the running conditions).

\section{The shield from environmental radioactivity}     \label{p:shield}

\begin{figure}[!htbp]
\centering
a) \includegraphics[width=7.0cm] {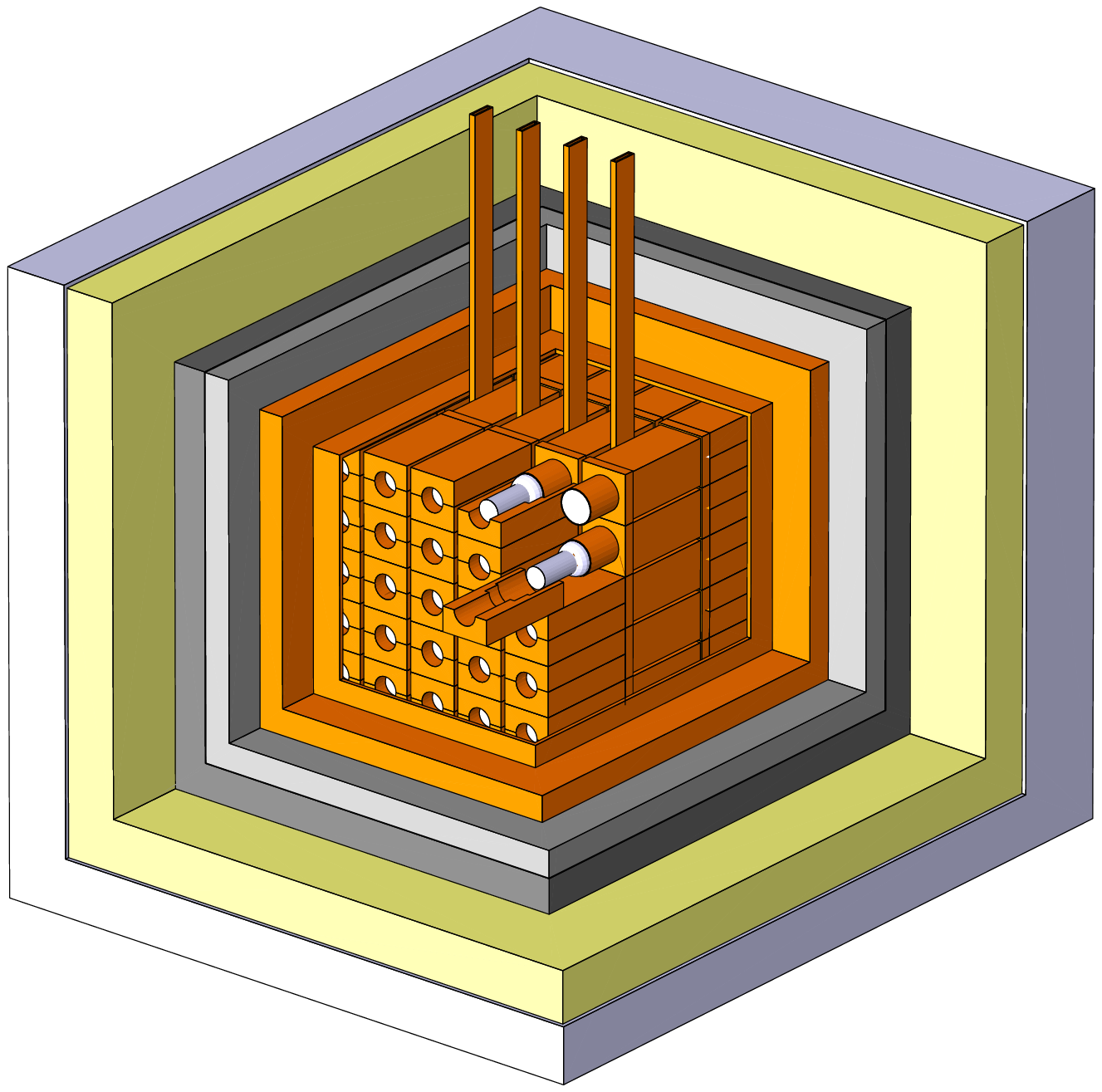}
b) \includegraphics[width=5.1cm] {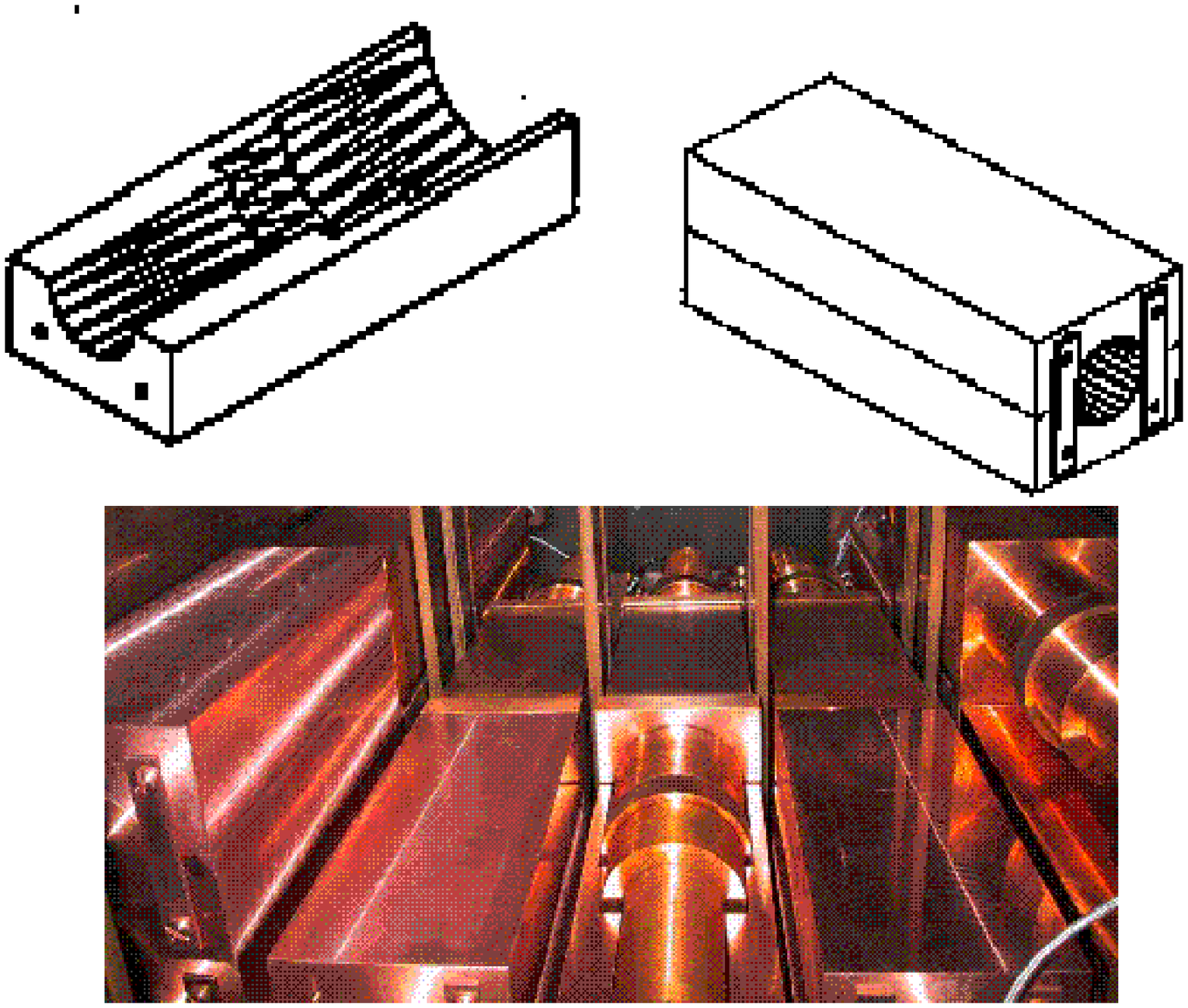}
\caption{a) A schematic view of the passive shield of the DAMA/LIBRA apparatus. Mostly outside the 
installation, the DAMA/LIBRA apparatus is also almost fully surrounded by about 1 m concrete made of
the Gran Sasso rock. 
b) {\it Up:} a scheme of the shaped low-radioactivity copper shield for the PMTs;
{\it Down:} detectors during installation; in the central and right up detectors the new 
shaped Cu shield surrounding light guides and PMTs was not yet applied. See text.}
\label{shield}
\end{figure}

Figure \ref{shield}a) shows a schematic view of the Cu/Pb/Cd/polyethylene-paraffin low background 
hard shield against environmental radioactivity made of very high radiopure materials, 
which 
are underground since at least about 15 years.
Moreover, paraffin/polyethylene fill the space surrounding the plexiglass box as much as possible depending on
the available space and, as mentioned, mostly outside the building, 
the DAMA/LIBRA apparatus is almost fully surrounded 
by about 1 m concrete, made from the Gran Sasso rock, which acts as a further neutron 
moderator\footnote{Neutron fluxes measured deep underground in the Gran Sasso National Laboratory
in the various energy regions are reported in refs. \cite{bel89,cri95,rin88,arn99}}.

\begin{figure}[!t]
\centering
a) \includegraphics[width=6.0cm] {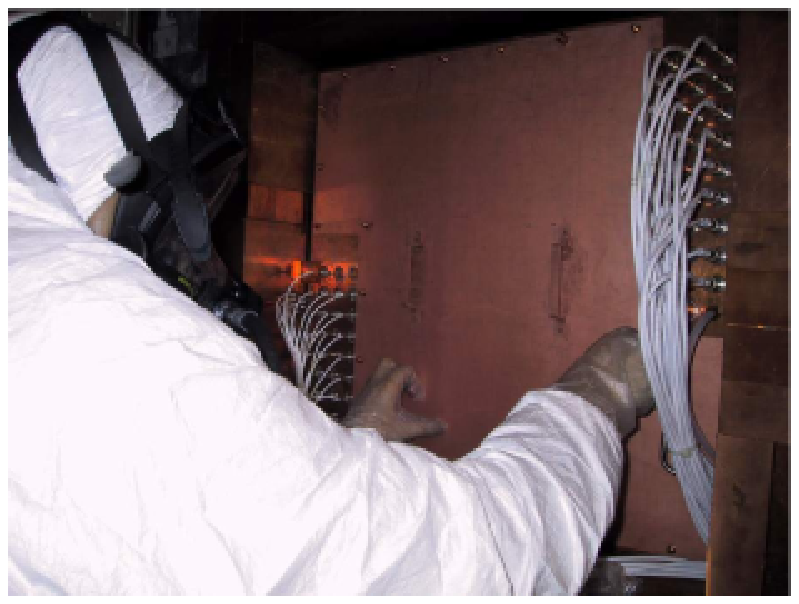}
b) \includegraphics[width=6.0cm] {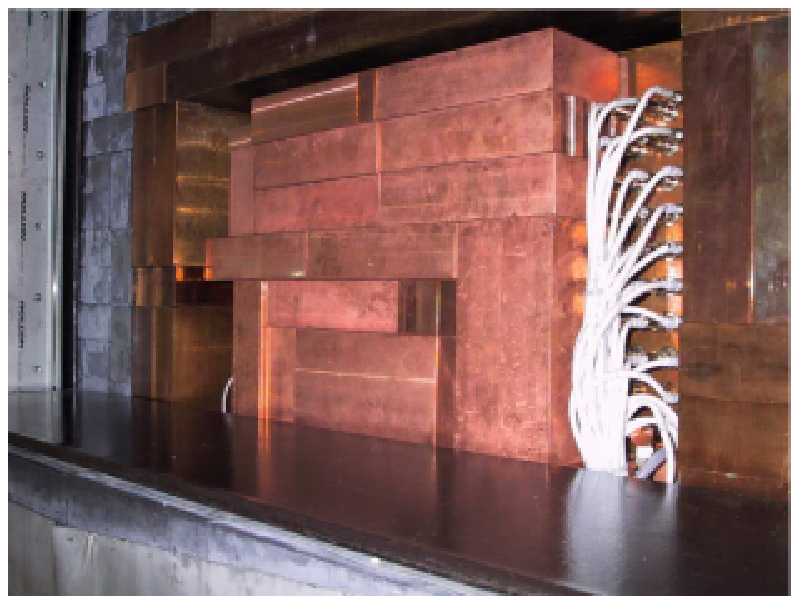}
\caption{a) Sealing the low radioactive OFHC Cu box. 
b) The space in front of the Cu box is filled by low radioactive Cu bricks and 
the front wall of the shield can go up.}
\vspace{-0.1cm}
\label{cubox}
\end{figure}

The 25 highly radiopure NaI(Tl) detectors are placed in 5 rows by 5 columns.
Each one has 9.70 kg mass and the size is: 
$10.2 \times 10.2 \times 25.4$ cm$^3$. They  constitute the sensitive part of the
DAMA/LIBRA apparatus and together with the PMTs (two for each detector at opposite sides) and 
their Cu shields are enclosed 
in a sealed low radioactive 
OFHC Cu box continuously flushed with HP N$_2$ longly stored deep underground (see Fig.~\ref{cubox}).
The Cu box is maintained at small overpressure with the respect to the environment, such as also 
the glove-box on the upper level (see Fig. \ref{setup}).

In the DAMA/LIBRA apparatus the PMTs and their 10 cm long light guides 
(made of Suprasil B \cite{supra} and directly coupled to the bare crystal acting 
also as optical windows) are surrounded by shaped 
low-radioactive OFHC freshly electrolyzed copper shields; see Fig.~\ref{shield}b).

All the materials constituting the Cu box and the Cu bricks, used to fill it as 
much as possible, have been selected for low radioactivity.
Particular care has been devoted to the long-living so-called standard contaminants:
$^{238}$U, $^{232}$Th and $^{40}$K, since the first two have very rich chains and the latter
one is present at a level of 0.0117\% in the $^{nat}$K, which is very abundant in nature.

The residual radioactivity in some components of the Cu box, that 
houses the NaI(Tl) detectors, is: 
i)   in the copper: $<$ 0.5 ppb, $<$ 1 ppb and $<$ 0.6 ppm of $^{238}$U, $^{232}$Th and $^{nat}$K, respectively;
ii)  in the feedthroughs: $<$ 1.6 ppb and $<$ 1.8 ppm of $^{232}$Th and $^{nat}$K, respectively;
iii) in the neoprene used in O-ring: $<$ 54 ppb and $<$ 89 ppm of $^{232}$Th and $^{nat}$K, respectively. 
All the limits -- at the level of sensitivity of the used Ge detector -- are at 95\% C.L.. 

Outside the Cu box, the passive shield is made by $\gsim$ 10 cm of OFHC low radioactive copper, 
15 cm of low radioactive lead, 1.5 mm of cadmium and about 10--40 cm of 
polyethylene/paraffin (thickness fixed by the available space); see Fig. \ref{shield}a).
The residual contaminants in some components of the passive shield are (95\% C.L.):
i)   in the copper see before;
ii)  in the Boliden Lead: $<$ 8 ppb, $<$ 0.03 ppb and $<$ 0.06 ppm of $^{238}$U, $^{232}$Th and $^{nat}$K, respectively;
iii) in the Boliden2 Lead: $<$ 3.6 ppb, $<$ 0.027 ppb and $<$ 0.06 ppm of $^{238}$U, $^{232}$Th and $^{nat}$K, respectively;
iv)  in the Polish Lead: $<$ 7.4 ppb, $<$ 0.042 ppb and $<$ 0.03 ppm of $^{238}$U, $^{232}$Th and $^{nat}$K, respectively;
v)   in the polyethylene: $<$ 0.3 ppb, $<$ 0.7 ppb and $<$ 2 ppm of $^{238}$U, $^{232}$Th and $^{nat}$K, respectively;
vi)  in the plexiglass: $<$ 0.64 ppb, $<$ 27.2 ppb and $<$ 3.3 ppm of $^{238}$U, $^{232}$Th and $^{nat}$K, respectively.

As mentioned, a sealed plexiglass box
(also continuously flushed with HP Nitrogen gas) encloses
the passive shield.
The Supronyl coverage (see above) and the mentioned about 1 m concrete complete the 
DAMA/LIBRA shield.
The sealed Cu box, the plexiglass box (both flushed with HP N$_2$) and the Supronyl represent the 
3-level sealing system from environmental air in the laboratory. 
Moreover, the Radon level inside the installation (that is just after the first more 
external level of sealing from external air) is continuously monitored and it is at the level of 
sensitivity of the Radon-meter; 
it is continuously recorded with the production data.

Specific procedures for preparing and handling the lead and copper bricks have also been 
selected. 
In particular, they have been chemically 
etched in a class 1000 clean room by using, respectively, HNO$_3$ and HCl aqueous solution in 
iper-pure water (conductivity = 0.0548 $\mu$S/cm at 25 $^{\circ}$C).
As an example, the main steps followed for the inner Cu etching are summarized here:
i)    Vessel I: pre-washing of the brick in iper-pure water;
ii)   Vessel II: washing in 1.5 l of super-pure HCl 3M;
iii)  Vessel III: first rinse with iper-pure water (bath);
iv)   Vessel IV: second rinse with iper-pure water (current);
v)    Vessel V: washing in 1.5 l of ultra-pure HCl 0.5M;
vi)   Vessel VI: first rinse with iper-pure water (bath);
vii)  Vessel VII: second rinse with iper-pure water (current);
viii) Vessel VIII: third rinse with iper-pure water (current);
ix)   The Cu brick is dried with selected clean towels and HP N$_2$ flux;
x)    Finally, the Cu brick is sealed in two envelopes (one inside the other) previously 
      flushed and filled with HP N$_2$ and stored underground until their mounting. 
The residual contaminants in the HCl used in solution with iper-pure water are certified by the producer. 
In particular, standard contaminants are quoted to be: 10 ppb of $^{nat}$K and 1 ppb of U/Th for super-pure 
HCl and 100 ppt of $^{nat}$K and 1 ppt of U/Th for ultra-pure HCl.
For each brick the bath was changed and after each step the solution of the bath was analyzed 
with ICP-MS technique. Residual contaminants were preliminarily checked in order to optimize the choice of 
the materials (in particular for gloves) and the cleaning procedure.
Moreover special tools have been built (by using teflon and high radiopure OFHC copper) to 
manage the bricks in order to minimize the contact with gloves in presence of acid solution.
ICP-MS measurements assure the same radiopurity characteristics of the iper-pure water 
used in the last rinse, before and after the rinse.

\section{The new highly radiopure NaI(Tl) detectors}

The $\simeq 250$ kg NaI(Tl) DAMA/LIBRA apparatus uses 25 NaI(Tl) highly radiopure 
detectors with 9.70 kg mass each one ($10.2 \times 10.2 \times 25.4$ cm$^3$ volume) placed in 
five rows by five columns. The granularity of the apparatus is an interesting feature for Dark 
Matter particle
investigations since Dark Matter particles can just contribute to events where only one of the 25 detectors
fires ($single-hit$ events) and not to whose where more than one detector fire in coincidence ($multiple-hit$ events).

The new DAMA/LIBRA detectors have been built by Saint Gobain Crystals and Detectors company. 
The constituting materials have been selected by using several techniques; moreover, chemical/physical 
purifications of the selected powders have been exploited. In addition, 
the more effective growing procedure has been applied and new rules for handling the bare 
crystals have been fixed.
Each detector is sealed in low radioactivity freshly electrolyzed copper housing and has two 10 cm 
long highly radiopure quartz (Suprasil B) light guides which also act as optical windows
being directly coupled to the bare crystals.

The bare crystals have been built by Kyropoulos method \cite{Kyr} in a platinum crucible according to 
a strict growing and handling protocol developed and agreed under confidential 
restriction\footnote{The confidential agreement protects the intellectual properties
both of the company and of the collaboration on the specific technical details.}. 
The seed, used in the growing, was from previous growth for highly radiopure NaI(Tl). 
The Kyropoulos method -- following the specific protocol agreed with the Saint Gobain 
Crystals and Detectors company -- has been adopted since it assures in particular: i)
possibility to operate in a well controlled overall situation for (radio)purity; 
ii) significant (radio)purification during the growing process; 
iii) high performances of the detectors. Other growing processes for NaI(Tl), 
such as the Bridgeman \cite{Kyr} was tested in the past 
and abandoned since it offered in our experience lower performances, 
mainly as regards the (radio)purification during growth.

The used NaI powders have preliminarily been selected for radiopurity by various methods; then,
chemical/physical procedures for the further
radiopurification have been exploited by Saint Gobain Crystals and Detectors company.
The final residual contaminants in the powders are
0.02 ppb for $^{238}$U, 0.02 ppb for $^{232}$Th and $<0.1$ ppm (95\% C.L.) for $^{nat}$K, respectively,
as measured by MS and AAS.
As regards the selected TlI powder, the residual contaminants in the 
powders after the chemical/physical purifications are 
0.8 ppb for $^{238}$U, 1.2 ppb for $^{232}$Th and $<0.06$ ppm (95\% C.L.) for $^{nat}$K, respectively,
as measured by MS and AAS.
It is worth noting that the concentration of Tl in the detectors is about 0.1\%.
Moreover, the growth with Kyropoulos method in
platinum crucible with a peculiar devoted protocol  
acts as a further relevant radiopurification step.
Obviously, the maximal sensitivity to determine the final residual contaminations 
is reached when directly measured in the crystals themselves (see later). In fact, the AAS and MS are mainly 
limited by the radiopurity of the line and by the added materials and -- as mentioned --
the used crystallization 
process itself acts as an additional considerable purification step. However, it is worth noting that 
the reachable 
final levels depend both on the initial impurity concentrations and on the used processes and
handling protocols. Particular care has been devoted to the selection of tools and abrasives to be used 
for cutting and polishing the crystals from the produced large ingots; this 
requires, in particular, severe selections and control of tools and of procedures 
in order to avoid any significant pollution on surface during the operation.

\begin{figure}[!b]
\centering
\includegraphics[width=10.cm] {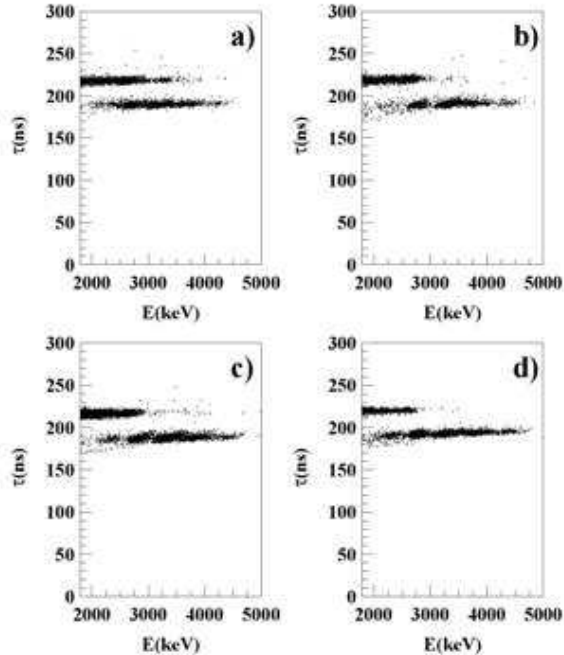}
\caption{First moment, $\tau$, of the time distribution for each event (recorded by the 
Waveform Analyser) -- calculated within a 600 ns averaging time -- as a function of the energy 
in unit of keV electron equivalent for four DAMA/LIBRA detectors.
The two populations $\gamma/$e and $\alpha$'s are clearly separated; the $\alpha$'s have shorter 
$\tau$'s.}
\label{a_discr}
\end{figure}

The number of materials entering in the production and in the assembling of a detector -- after the
ingots production -- has been minimized and deeply selected for
radiopurity. In particular selected Tetratec-teflon foils, wrapping the bare cystals, 
have been used as diffusers of the scintillation light 
because of their suitable performances and mainly of their radiopurity with respect to 
other possibilities.
For the same reason, OFHC freshly electrolyzed copper has been
used for the housings of the detectors. 
Preliminary suitable etching of the copper housings has been performed
for a deep cleaning of possible residuals from machinery. Obviously, 
the foam (not distinguishable from outside by eyes), usually used inside the housing of standard
NaI(Tl) detectors for compensation, is not used in the construction of the DAMA/LIBRA detectors
because of the low background requirements. Instead, a mechanical compensation 
for the different specific heats of the Cu and NaI(Tl) 
is included in the design of the copper housing to avoid any possible damages during transportation 
in non-air-conditioned environment before installation deep underground.
Many other specific expedients have also been exploited.
In particular we remind that the detectors, built for the DAMA apparata in Gran Sasso, 
have always been cut and assembled just after the
growth of the crystalline bulk in a glove-box in controlled atmosphere, tightly
sealed according to special requirements, road transported and immediately brought underground.
 

A preliminary estimate of the residual U and Th contamination in a detector can be obtained from the internal 
$\alpha$ particles produced by both chains. They can be measured by exploiting the $\alpha/$e pulse shape 
discrimination in NaI(Tl), which has practically 100\% effectiveness in the MeV range.
Fig. \ref{a_discr} shows some examples of the distribution of $\tau$ as a function of energy
for four DAMA/LIBRA detectors.
The variable $\tau = \frac{\Sigma_i h_i t_i}{\Sigma_i h_i }$ is the first moment of the time distribution
of each scintillation pulse (as recorded
by the Waveform Analyser; see Sect. \ref{sec_ele}); 
there $h_i$ is the pulse height at the time 
$t_i$ and the sum is over 600 ns after the starting of the 
pulse. The two populations $\gamma/$e and $\alpha$'s are clearly separated; the $\alpha$ particles have 
shorter $\tau$ values.
The measured $\alpha$ yield in the DAMA/LIBRA detectors ranges from 7 to some tens $\alpha$/kg/day.

\vspace{0.2cm} 
\noindent {\bf $^{232}$Th}
\vspace{0.1cm}

The time-amplitude method\footnote{The technique of the time-amplitude analysis is based on 
the identification of radioactive isotopes by means of the characteristic time sequence
of peculiar events ($\alpha$, $\beta$, $\gamma$ of known energy) 
given by the subsequent decays of 
daughter nuclides; see e.g. ref.
\cite{tam}.} has been used 
to determine the activity of the $^{228}$Th (T$_{1/2} = 1.912$ yr) 
subchain in the $^{232}$Th family.
In fact, the arrival time and energy of the events in each crystal can be used to identify a 
sequence of fast $\alpha$ decays following $^{228}$Th decay: 
$^{224}$Ra (Q$_{\alpha}$ = 5.789 MeV, T$_{1/2}$ = 3.66 d) $\rightarrow$ $^{220}$Rn 
(Q$_{\alpha}$ = 6.405 MeV, T$_{1/2}$ = 55.6 s) $\rightarrow$ $^{216}$Po (Q$_{\alpha}$ = 6.906 MeV, 
T$_{1/2}$ = 0.145 s) $\rightarrow$ $^{212}$Pb.
\begin{figure}[!ht]
\centering
\includegraphics[width=12.cm] {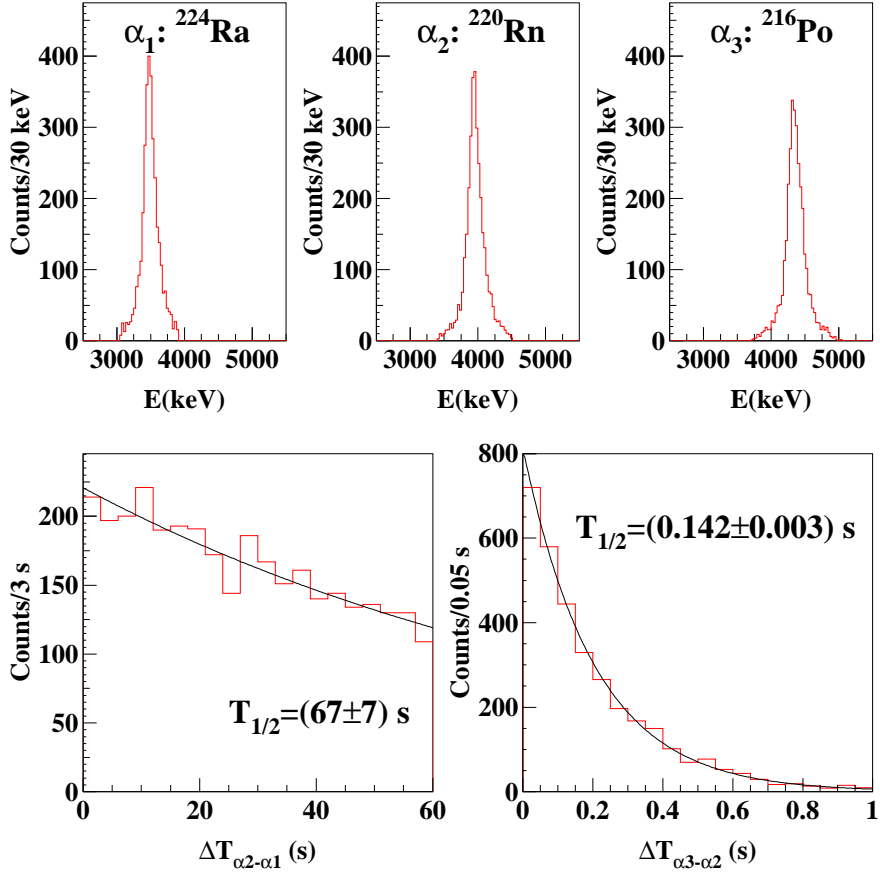}
\vspace{-0.35cm}
\caption{{\it Top:} $\alpha$ peaks of $^{224}$Ra, $^{220}$Rn and $^{216}$Po selected by the time-amplitude 
analysis (see text) from the data accumulated during an exposure of 8100 kg $\times$ day for the triple
delayed coincidence with the detector $d)$ of Fig. \ref{a_discr}.
The energy resolutions ($\sigma$) are: 75, 83, and 90 keV, well compatible with those expected
by $\gamma$'s calibrations (see Sect. \ref{sec_gf}).
{\it Bottom:} distributions of the time intervals between events due to $^{224}$Ra and $^{220}$Rn $\alpha$ 
decays ($left$) and between events due to $^{220}$Rn and $^{216}$Po $\alpha$ decays ($right$).
The obtained half-lives of $^{220}$Rn: $(67 \pm 7)$ s, and of $^{216}$Po: $(0.142 \pm 0.003)$ s, are in agreement 
with the table values (55.6 s and 0.145 s, respectively \cite{toi98}).}
\vspace{-0.1cm}
\label{timamp}
\end{figure}

An example of the obtained results for the detector $d)$ of Fig. \ref{a_discr} is shown in Fig. \ref{timamp}.
The subchain has been selected by looking for events 
in the energy window 3050-3900 keV, followed within 60 s by another event
in the energy window 3400-4500 keV, still followed within 1 s by the last event
in the energy window 3650-5100 keV.

The $\alpha$ peaks as well as the 
distributions of the time intervals between the events are in a good agreement with those expected for the 
$\alpha$'s from the $^{224}$Ra $\rightarrow$ $^{220}$Rn $\rightarrow$ $^{216}$Po $\rightarrow$ $^{212}$Pb 
chain. 

Moreover, the position of the $\alpha$ peaks, reported on an energy scale calibrated with $\gamma$ sources, 
allows the determinination of the $\alpha/\beta$ light ratio in the used crystal; it is:
$\alpha/\beta = 0.467(6)+0.0257(10) \times E_{\alpha}[MeV]$, where $E_{\alpha}[MeV]$ is the energy of the 
$\alpha$ particle in MeV.

The number of the triple delayed coincidences, selected by the given procedure, is 3310
in an exposure of 8100 kg $\times$ day.
It corresponds, after applying the efficiencies, to $(9.0\pm0.4)$ $\mu$Bq/kg.

Repeating the same analysis for all the crystals of the DAMA/LIBRA apparatus we obtain a 
$^{228}$Th activity
ranging from 2 to about 30 $\mu$Bq/kg depending on the crystal.

For completeness we note that the activity value determined 
in this way for each detector is well consistent with that obtained in an independent 
analysis of the so-called Bi-Po events. The Bi-Po events are due to the $^{212}$Bi $\beta$ decay to 
$^{212}$Po and to the subsequent $^{212}$Po $\alpha$ ($Q_{\alpha}=8.954$ MeV) decay to $^{208}$Pb with $T_{1/2} 
= 299$ ns 
\footnote{Also the $^{238}$U radioactive chain produces Bi-Po events, due to $^{214}$Bi $\beta$ decay to $^{214}$Po and 
to the subsequent $^{214}$Po $\alpha$ ($Q_{\alpha}=7.833$ MeV) decay to $^{210}$Pb with $T_{1/2}=164.3 \mu$s. 
In particular, their rate is connected with the activity of the $^{226}$Ra $\rightarrow$ $^{210}$Pb decay chain segment.}.
They can be clearly identified by the recorded pulse shapes over a 2048 ns time window.
In particular, for the detector $d)$ of Fig. \ref{a_discr} the inferred activity for $^{228}$Th decay subchain is
$(9.4\pm1.5)$ $\mu$Bq/kg, well compatible with the previous determination. 

If the whole $^{232}$Th radioactive chain is assumed at equilibrium (as it can be the case considering that 
the longest half-life of the daughter isotopes, $^{228}$Ra, is 5.75 years), the $^{232}$Th 
contents in the crystals typically range from 0.5 ppt to 7.5 ppt.

\vspace{0.9cm} 
\noindent {\bf $^{238}$U}
\vspace{0.1cm}

An estimate of the $^{238}$U contamination can be obtained for each detector from the measured $\alpha$ 
activity and the $^{232}$Th content evaluated above. 
Assuming -- as a first approximation -- that also the $^{238}$U chain is in equilibrium,
one obtains values ranging from 0.7 to 10 ppt depending on the crystal.

\begin{figure}[!htbp]
\vspace{-0.25cm}
\centering
\includegraphics[width=9.cm]{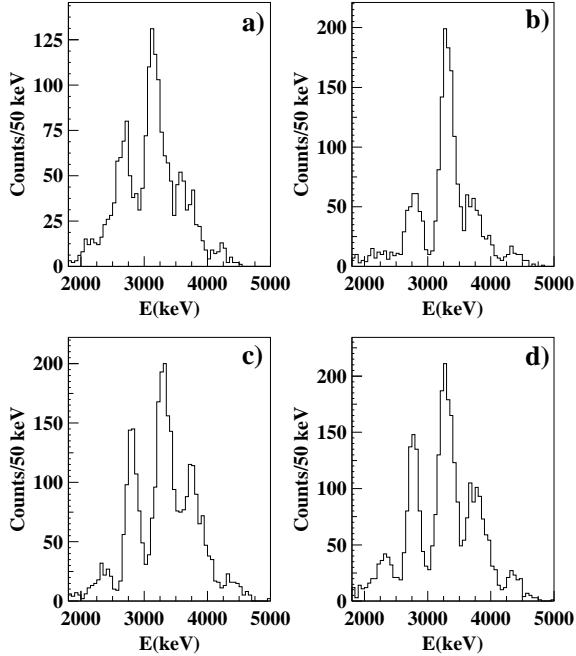}
\vspace{-0.35cm}
\caption{The $\alpha$ energy distributions in some of the NaI(Tl) crystals corresponding to 
a live time of 570 hours. 
The energy is given in keV electron equivalent.}
\vspace{-0.1cm}
\label{a_spectra}
\end{figure}

However, the hypothesis of equilibrium for the $^{238}$U chain in the detectors 
is not confirmed by the study of the energy 
distributions of the $\alpha$ particles, which can allow in principle the determination of the various 
contributions 
from the $^{238}$U subchains.
In Fig.~\ref{a_spectra} we show -- as an example -- the distributions of $\alpha$ from $^{238}$U 
and $^{232}$Th chains in some 
of the new NaI(Tl) crystals as collected in a live time of 570 hours; 
there the $\alpha$ energies are given in keV electron equivalent.

In particular, starting from the low energy peak, the five $\alpha$ peaks in the energy spectra of 
Fig.~\ref{a_spectra} can be associated with\footnote{It is worth noting that the $\alpha$
associated with the decays of $^{212}$Po and of $^{214}$Po are not present in the shown $\alpha$ plots
because they belong to a Bi-Po event and they are mainly vetoed by the acquisition system (see later).}: 
  i) $^{232}$Th ($Q_{\alpha}=4.08$ MeV) + $^{238}$U  (4.27 MeV); 
 ii) $^{234}$U  (4.86 MeV) + $^{230}$Th (4.77 MeV) + $^{226}$Ra (4.87 MeV); 
iii) $^{210}$Po (5.41 MeV) + $^{228}$Th (5.52 MeV) + $^{222}$Rn (5.59 MeV) + $^{224}$Ra (5.79 MeV); 
 iv) $^{218}$Po (6.12 MeV) + $^{212}$Bi (6.21 MeV) + $^{220}$Rn (6.41 MeV) and
  v) $^{216}$Po (6.91 MeV). 
Thus, the contribution of each $\alpha$ decay has been simulated and fitted to the experimental energy 
spectra (some examples are given in Fig.~\ref{a_spectra})
considering the $^{238}$U radioactive chain split into five segments ($^{238}$U $\rightarrow$ 
$^{234}$U $\rightarrow$ $^{230}$Th $\rightarrow$ $^{226}$Ra $\rightarrow$ $^{210}$Pb $\rightarrow$ 
$^{206}$Pb) and the $^{232}$Th chain at equilibrium.

The fit of the measured alpha spectra allows the determination of the activities of the five $^{238}$U 
subchains and of the $^{232}$Th chain. The results confirm the hypotheses that 
the $^{238}$U chain is broken in these NaI(Tl) crystals. 
As an example in the detector $d)$ of Fig.~\ref{a_discr} and Fig.~\ref{a_spectra}
the $^{232}$Th and $^{238}$U contents obtained by the fit are:
  1) $(8.5 \pm 0.5)$ $\mu$Bq/kg of $^{232}$Th [that is, $(2.1 \pm 0.1)$ ppt, value in agreement with the two 
     determinations given above using the time-amplitude and the Bi-Po analyses];
  2) $(4.4 \pm 0.7)$ $\mu$Bq/kg for $^{238}$U $\rightarrow$ $^{234}$U decay subchain [that is, $(0.35 \pm 0.06)$ ppt 
     of $^{238}$U];
  3) $(15.8 \pm 1.6)$ $\mu$Bq/kg for $^{234}$U $\rightarrow$ $^{230}$Th + $^{230}$Th $\rightarrow$ $^{226}$Ra decay 
     subchains (they all contribute to the same peak);
  4) $(21.7 \pm 1.1)$ $\mu$Bq/kg for $^{226}$Ra $\rightarrow$ $^{210}$Pb decay subchain and
  5) $(24.2 \pm 1.6)$ $\mu$Bq/kg for $^{210}$Pb $\rightarrow$ $^{206}$Pb decay subchain.

As it is clear e.g. from Fig.~\ref{a_spectra}, the residual contaminants may
be slightly different even among detectors made from NaI(Tl) crystals grown with the same
selection of materials, purification processes and protocols.
In fact, some casual pollutions during the growth and handling procedures may in principle be possible, being 
the detectors built in an industrial environment. Differences may also arise depending 
on the use of different bulks or on which part
of a crystallized bulk has been used to build 
the detector. In fact, the purification during crystallization may be not uniform in the whole bulk mass. 
Moreover, the uniformity of the contaminants 
distribution inside the total material needed to construct each part of the detectors cannot be 
assured. 
Obviously, casual pollution may also occur when handling the detectors in industrial environment 
or deep underground without the needed extreme care.

\vspace{0.1cm} 
\noindent {\bf $^{nat}$K}
\vspace{0.1cm}

An estimate of the potassium content in the DAMA/LIBRA crystals has been obtained investigating over large 
exposure the presence of peculiar double coincidences.
In fact, the $^{40}$K (0.0117\% of $^{nat}$K)
also decays by EC to the 1461 keV level of $^{40}$Ar (b.r. 10.66\%) 
followed by X-rays/Auger electrons, that are contained in the crystal with efficiency $\sim 1$, and a 1461 keV 
de-excitation $\gamma$.
The latter one can escape from one detector (hereafter $A$) and hit another one, causing the double 
coincidence.  The X-rays/Auger electrons give rise in the detector $A$ to a 3.2 keV peak, binding 
energy of shell K in $^{40}$Ar\footnote{In the $(76.3 \pm 0.2)\%$ of the cases an electron from shell K 
(E$_K = 3.2$ keV) is involved in the process, in the $(20.9 \pm 0.1)\%$
an electron from shell L (E$_L = 0.3$ keV) 
and in $(2.74 \pm 0.02)\%$ electron from upper shells.}.

\begin{figure}[!ht]
\centering
\includegraphics[width=8.cm]{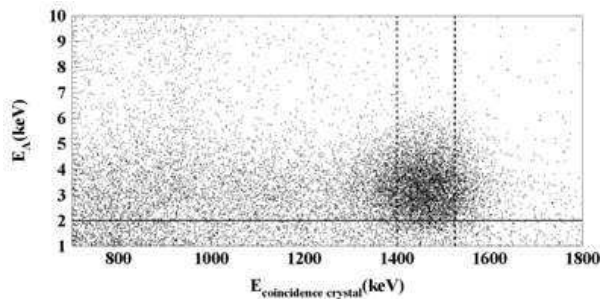}
\vspace{-0.35cm}
\caption{Example of the analysis to determine $^{nat}$K contamination in one of the 25 crystals (see text).
The scatter plot shows the low energy region of the considered crystal, $A$, as a function of the energy 
detected in the other crystal involved in the double coincidence. 
The threshold of each PMT is at single photoelectron level. For comparison, the software energy threshold
used in the data analyses of the {\it single-hit} events for Dark Matter particle investigation:
2 keV, is shown as continuous line.}
\vspace{-0.2cm}
\label{k40_8}
\end{figure}

The experimental data have been analyzed searching for these double coincidences;
Fig.~\ref{k40_8} shows as an example a scatter plot of the energies of the detector $A$ and all 
the other detectors involved in the coincidence.
It is evident a spot that correlates the 1461 keV events in the other crystals with the 3.2 keV peak 
in crystal $A$. 
The detection efficiency for such coincidences has been evaluated for each crystal by  MonteCarlo code.
The analysis has given for the $^{nat}$K content in the crystals values not exceeding about 20 ppb.
It is worth noting that the identification of the 3.2 keV peak offers also an intrinsic calibration 
of each DAMA/LIBRA detector in the lowest energy region (see Sect. \ref{sec_gf}).

\vspace{0.1cm} 
\noindent {\bf $^{125}$I}
\vspace{0.1cm}

As regards possible non-standard contaminants, we remind that during the period of powders' 
storage and/or crystal growing and handling at sea level (of order of at least some months 
for large crystals) radioactive isotopes can be produced by cosmic-ray interactions.
Therefore, it is good practice in this kind of experiments to wait for the decay of the 
short-life cosmogenic isotopes before considering the production data in underground.

\begin{figure}[!htbp]
\centering
\includegraphics[width=10.0cm] {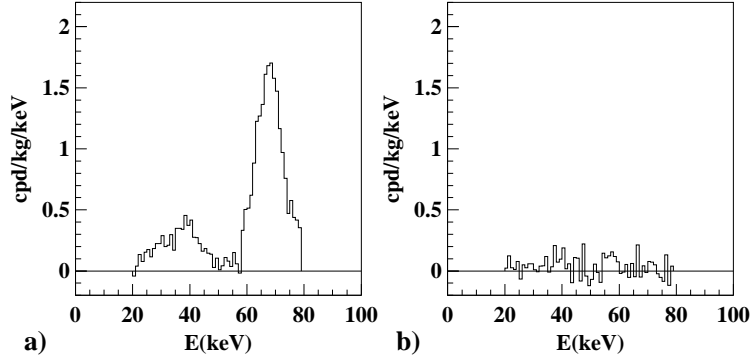}
\vspace{-0.1cm}
\caption{a) Difference between the energy distributions collected at Gran Sasso with a DAMA/LIBRA NaI(Tl) detector 
just stored underground and $~15$ months later. The role of $\simeq$ 67 keV peak due to $^{125}$I 
activation (half-life 59.40 days) is evident.
In particular, the EC of $^{125}$I from shell K (80.12\%) gives rise to the 67.3 keV peak, while EC 
from upper shells produces the $\simeq 40$ keV peak.
b) The same as a) but obtained for a detector stored underground for a longer time; the peaks due to 
$^{125}$I are not present.}
\vspace{-0.1cm}
\label{fig70}
\end{figure}

In particular, here we comment the case of cosmogenic $^{125}$I which decays by EC to $^{125}$Te giving 
35.5 keV $\gamma$'s (or internal conversion electrons) 
and, at the same time, tellurium X-rays and Auger electrons, whose total 
energy is 31.8 keV, giving rise to a peak around $\simeq$ 67 keV with half-life 59.40 days 
(see Fig.~\ref{fig70}).
In this case the best situation is reached after $\simeq$ 8 months of storage deep underground. 

Some arguments on cosmogenic activation in NaI(Tl) can be found in ref.~\cite{dye91}, while the production 
rate of cosmogenic isotopes can be calculated mainly according to the model of ref.~\cite{sil73}.
However, the activity of cosmogenic isotopes in NaI(Tl) detectors would be in practice 
generally lower than the calculated one because of purification and saturation processes 
during the growing procedures.

\vspace{0.1cm} 
\noindent {\bf $^{129}$I and $^{210}$Pb}
\vspace{0.1cm}

\begin{figure}[!htbp]
\centering
\includegraphics[width=6.0cm] {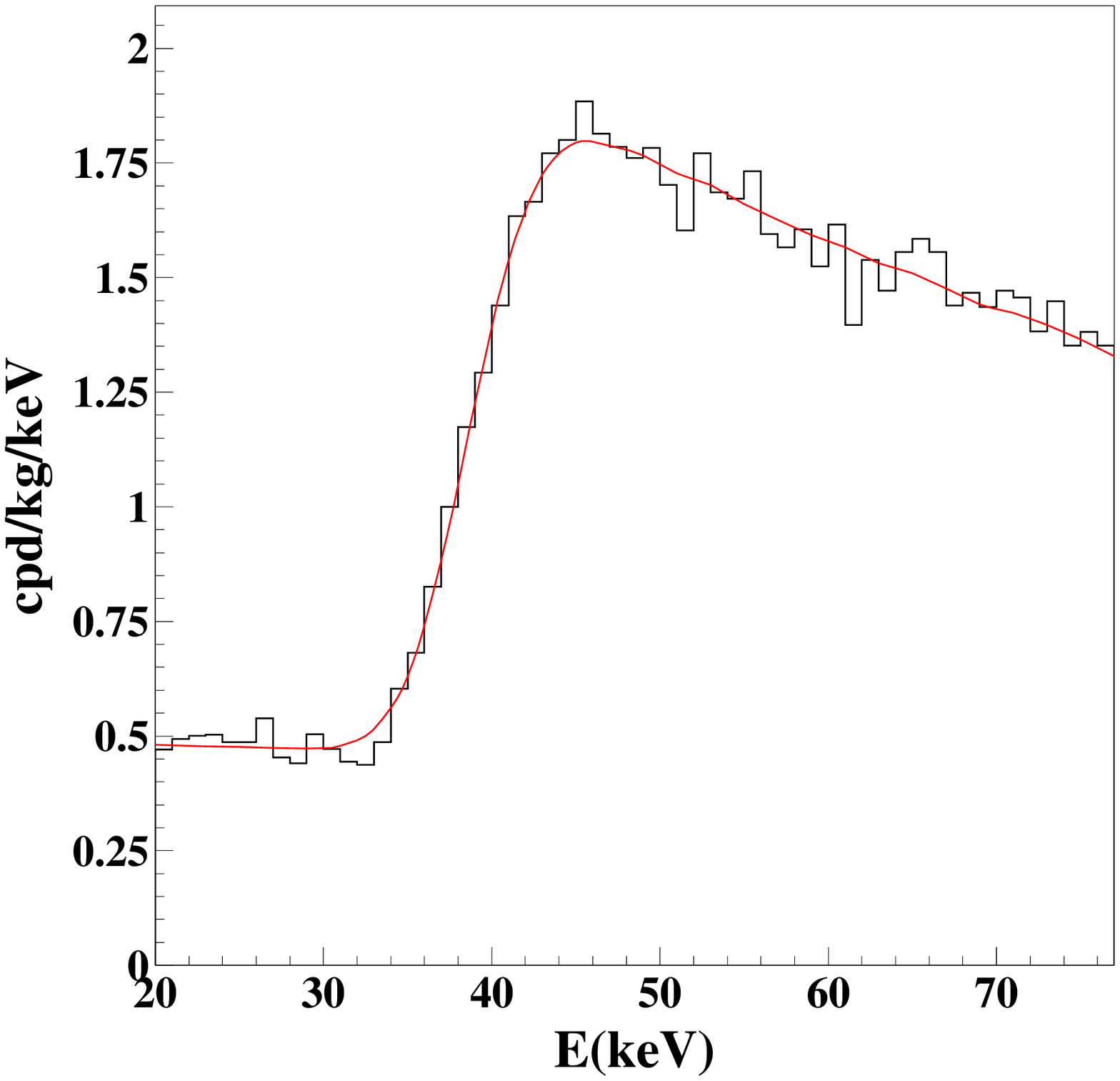}
\includegraphics[width=6.0cm] {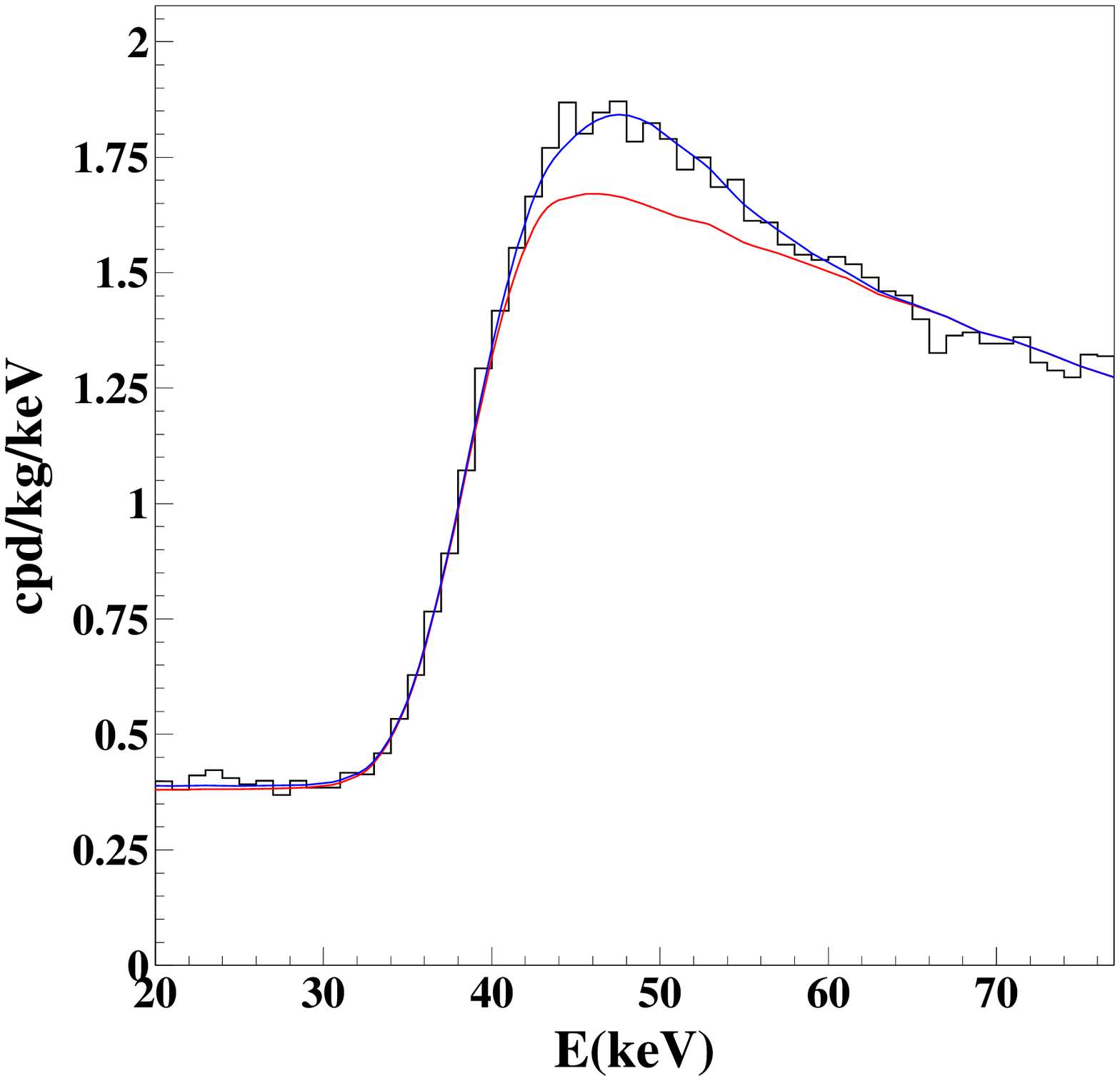}
\vspace{-0.1cm}
\caption{Energy distributions of two new DAMA/LIBRA NaI(Tl) detectors in the energy region of interest 
for the cosmogenic $^{129}$I decay process (histogram) and the model of background (lines).
The contribution of internal $^{210}$Pb is visible only in the plot of the second crystal
($<12$ $\mu$Bq/kg at 90\% C.L. and $(29 \pm 3)$ $\mu$Bq/kg, respectively).
The amount of cosmogenic $^{129}$I has been estimated to be 
$^{129}\textrm{I}/^{nat}\textrm{I} = (1.7\pm0.1)\times 10^{-13}$ for both the crystals.}
\vspace{-0.1cm}
\label{fg:129i}
\end{figure}

The cosmogenic $^{129}$I (T$_{1/2} = 1.57 \times 10^{7}$ yr) can be present in the natural Iodine
with a percentage of the order of $1.5 \times 10^{-12}$ (considering sediments below the zone of
bioturbation, i.e. without the presence of anthropogenic Iodine  \cite{iod129}). It decays by $\beta^-$ 
into the excited level of $^{129}$Xe at 39.57 keV. Thus, this process produces a de-excitation 
$\gamma$ of 39.57 keV (or internal conversion electron) plus an electron 
with energy distribution according to the beta spectrum of $^{129}$I (maximum energy 154 keV).
Therefore, the total energy deposited in the crystal from this process has a peculiar distribution with 
sharp rise around $\simeq$ 40 keV and smooth behaviour up to $\simeq$ 194 keV (sum of the gamma energy 
and of the maximum energy of the electrons). 
Fig. \ref{fg:129i} shows the energy spectra of
two detectors in the energy region of interest for this process and the model of background.
Hence, the amount of cosmogenic $^{129}$I has been estimated to be there 
$^{129}\textrm{I}/^{nat}\textrm{I} = (1.7\pm0.1)\times 10^{-13}$.
If this value is used for dating the NaI powders, one obtains that they have been extracted from an ore 
with an age of order of 50 Myr.

Fig. \ref{fg:129i} also allows an independent evaluation of the content of $^{210}$Pb in the crystals. 
The $^{210}$Pb isotope $\beta$ decays with a branching ratio of 84\% into the excited level 
of $^{210}$Bi at 46.5 keV; the electron has an energy distribution according to the beta spectrum
(maximum energy: 16.6 keV).
The presence of a contamination of $^{210}$Pb mainly on the housings' surfaces has been used by the DAMA/NaI 
apparatus as
a calibration point at 46.5 keV (due to the gamma from the de-excitation of the $^{210}$Bi excited level) during 
its data taking (see for example \cite{Nim98,RNC}).
Such a peak is not so appreciable in the new DAMA/LIBRA detectors;
anyhow, the contents of internal $^{210}$Pb can be determined 
for these detectors by investigating the energy region around $\simeq$ 50 keV. For the two 
detectors of Fig. \ref{fg:129i} one gets: $(7 \pm 4)$ $\mu$Bq/kg
(safely corresponding to an upper limit of $<12$ $\mu$Bq/kg at 90\% C.L.) and $(29 \pm 3)$ $\mu$Bq/kg,
respectively. Let us note that for the second detector in Fig. \ref{fg:129i} -- corresponding to
the detector $d)$ of Fig.~\ref{a_discr} and Fig.~\ref{a_spectra} -- an independent determination of 
$^{210}$Pb content has been mentioned above when studying the $\alpha$ spectrum in the crystals:
the two values are well compatible.

Repeating the same analysis for all the new DAMA/LIBRA detectors, we obtain that 
the amount of cosmogenic $^{129}$I is at the same level for all the new DAMA/LIBRA detectors;
in addition, the content of $^{210}$Pb in the crystals typically ranges: $(5 - 30)$ $\mu$Bq/kg.

\vspace{1.0cm} 
\noindent {\bf $^{22}$Na}
\vspace{0.1cm}

Among the possible isotopes produced by cosmogenic activation at sea level in NaI(Tl) 
we also investigate the $^{22}$Na 
(T$_{1/2}$ = 2.6 yr). The calculated maximum rate level is $\simeq$ 100 cpd/kg at sea level.
An estimate of the $^{22}$Na activity in the detectors can be obtained by searching for triple 
coincidences induced by the $\beta^+$ decay of $^{22}$Na followed by 1274.6 keV de-excitation $\gamma$ 
(b.r. 90.33\%).
In particular, we have searched for events where the positron and one of the two 511 keV annihilation 
$\gamma$ release all their energy in one detector while the second 511 keV annihilation $\gamma$ and the 1274.6 keV 
$\gamma$ hit other two detectors. For example, choosing an energy window of 650-1000 keV for the first detector
and $\pm 2 \sigma$ around 511 keV for the second one, we look for the typical peak at 1274.6 keV 
in the third detector.
Accounting for the detection efficiency of the process, the obtained activity in the detectors ranges from 
an upper limit of $15$ $\mu$Bq/kg (90\% C.L.) to some tens $\mu$Bq/kg depending 
on their arrival time underground.

\vspace{0.6cm} 
\noindent {\bf $^{24}$Na}
\vspace{0.1cm}

Another source of internal background is the possible presence of $^{24}$Na.

In fact, environmental neutrons would induce the reaction $^{23}$Na(n,$\gamma$)$^{24}$Na with 
0.1 barn cross-section and the reaction $^{23}$Na(n,$\gamma$)$^{24m}$Na with 0.43 barn 
cross-section \cite{toi78}.
The $^{24}$Na isotope is a $\beta$-emitter (end point equal to 1.391 MeV) with two prompt
associated $\gamma$'s (2.754 and 1.369 MeV). On the other hand, the $^{24m}$Na isotope decays 
100\% of the times in $^{24}$Na by internal transition with a $\gamma$ of 0.472 MeV.
Thus, the presence of $^{24}$Na has been investigated with high sensitivity 
by looking for triple coincidences induced by
a $\beta$ (E1) in one detector and the two $\gamma$'s in two adjacent ones (E2, E3). 
In particular, we consider: 
i) 0.57 MeV $<$ E1 $<$ 1.3 MeV, 
ii) E2 and E3 inside $\pm 1 \sigma$ energy windows around the photopeak positions.
Only one event satisfying the requirements has been found in the considered data set 
(106376 kg$\times$day); it can also be ascribed to side processes (for example
to the decay of $^{208}$Tl in the $^{232}$Th chain). 
Safely we consider that the number of detected events is lower than 3.89 (90\% C.L.) following the 
procedure given in \cite{hag02}; thus, accounting for the detection efficiency of the process, 
an upper limit of 2409 $^{24}$Na decays (90\% C.L.) is obtained for the considered exposure.
It corresponds to an activity $< 0.26$ $\mu$Bq/kg (90\% C.L.).
Assuming the production of these short-life isotopes as fully due to a steady thermal 
neutron flux, an upper limit on the thermal neutron flux surviving 
the multicomponent DAMA/LIBRA shield can be derived as: $< 1.2 \times 10^{-7}$ cm$^{-2}$s$^{-1}$ (90\% C.L.).

\vspace{1.cm} 
\noindent {\bf Miscellanea}
\vspace{0.1cm}

Finally, as regards the $^{3}$H (T$_{1/2}$ = 12.3 yr)\footnote{The $^{3}$H can be produced by spallation processes induced 
by cosmic rays on $^{23}$Na and $^{127}$I. 
In particular, we recall: $n$ + $^{23}$Na $\rightarrow$ $^{3}$H + $^{21}$Ne(stable) and $n$ + 
$^{127}$I $\rightarrow$ $^{3}$H + $^{125}$Te(stable), for neutron with energy $\gsim$ 15, 20 MeV.},
the calculated maximum rate is $\simeq$ 20 cpd/kg at sea level.
However, as mentioned above, in real conditions the $^{3}$H activity in a NaI(Tl) detector 
would be essentially lower than the calculated maximum level, because it would be extracted 
in the process of NaI(Tl) purification and growth. 
In our crystals the $^{3}$H content has been estimated  to be: $< 0.9 \times 10^{-4}$ Bq/kg at 95\% C.L..
Moreover, for the $^{87}$Rb (T$_{1/2} = 4.8 \times 10^{10}$ yr), and for the $^{85}$Kr (T$_{1/2}$ = 10.7 yr)
isotopes we just mention the obtained upper limits (95\% C.L.): $< 3 \times 10^{-4}$ Bq/kg
and $< 10^{-5}$ Bq/kg, respectively.

\section{The photomultipliers}

The photomultipliers used in this experiment have been built by Electron Tubes Limited 
and longly stored underground.
The PMTs are made of ultra-low background glass. 
The materials entering in the construction of the PMT have been selected by various kind of measurements
and, in particular, measurements on samples have also been carried out with low background DAMA/Ge detector 
deep underground. 

\begin{figure}[!h]
\vspace{-0.2cm}
\centering
\includegraphics[width=8.5cm] {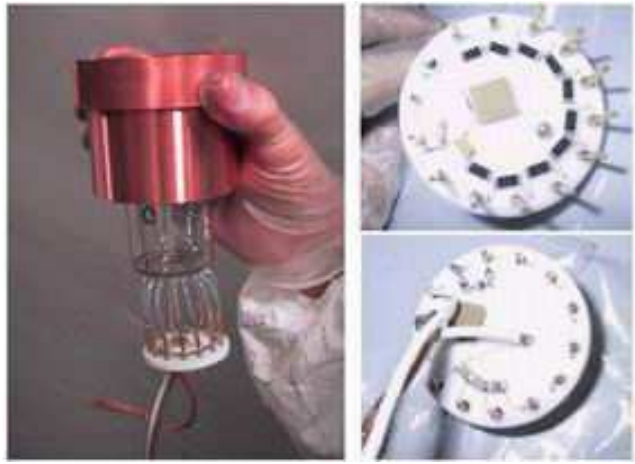}
\vspace{-0.2cm}
\caption{{\it Left:} a DAMA/LIBRA PMT with flying leads directly connected 
to the voltage divider.
{\it Right:} a voltage divider assembled with miniaturized SMD resistors and capacitors mounted on 
thin teflon socket.}
\label{pmtfoto}
\end{figure}

The PMTs have flying leads (instead of a hard socket) and are directly connected 
to suitable voltage dividers (see Fig.~\ref{pmtfoto}). 
The voltage dividers have been assembled deep underground with miniaturized SMD resistors and capacitors 
(also selected for radiopurity by low background Ge detector) and mounted on thin 
teflon sockets (see Fig.~\ref{pmtfoto}).
All the solders were performed by using low radioactive Boliden lead and low radioactive resin.

\begin{figure}[!htbp]
\vspace{-0.3cm}
\centering
\includegraphics[width=\textwidth] {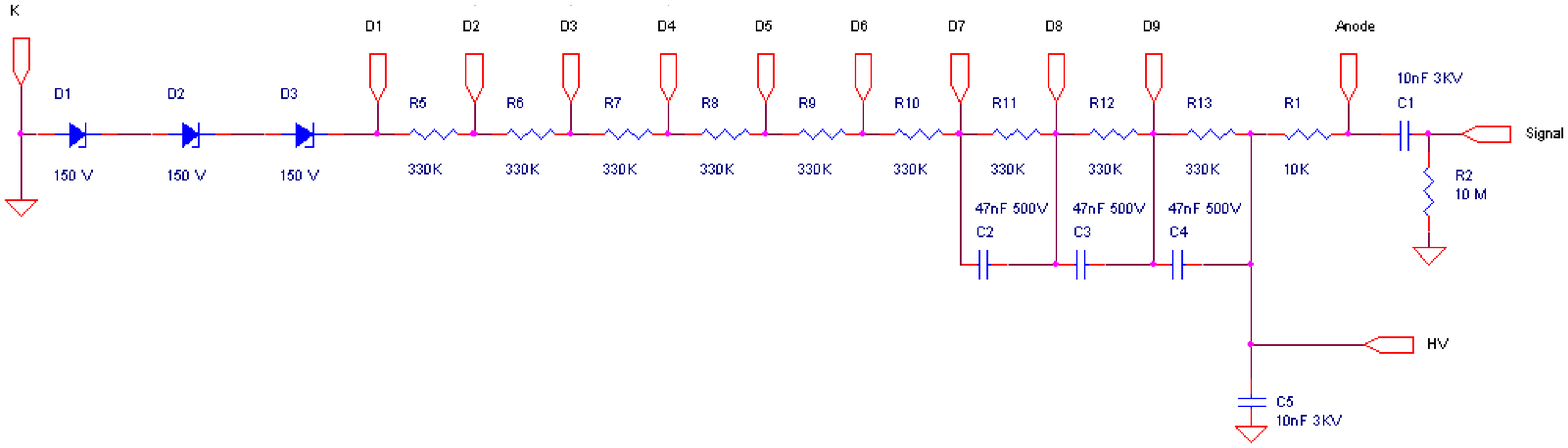}
\vspace{-0.6cm}
\caption{The voltage divider scheme.}
\label{pmtVD}
\end{figure}

The voltage dividers have been optimized to reach the best signal/noise ratio; a scheme 
is shown in Fig. \ref{pmtVD}. 

The applied High Voltage is positive; in this way 
the cathode is mantained to ground in order to avoid any voltage difference with the PMT window,
light guide and the copper housing.

The PMTs work near the maximum high voltage -- within the linearity plateau -- in order to 
assure a suitable energy threshold and to improve the PMT noise rejection near it.

The PMTs have nine high gain, high stability dynodes of linear focused design, high 
quantum efficiency ($\simeq 30$\% at 380 nm), good pulse height resolution for single 
photoelectron pulses (peak/valley $\gsim 2$), low dark noise rate ($\simeq 0.1$ kHz), 
and a gain of $\simeq 10^6$.
In the experiment each detector has two PMTs both to increase the light collection and to 
reduce the PMT noise contribution near energy threshold. They work in coincidence at single photoelectron 
threshold.

The residual contaminants in the PMTs have been measured with and without a bent light guide 
(i.e. shielding and not shielding the PMT contribution) and a low background NaI(Tl); typically
$^{232}$Th and $^{238}$U are lower than few tens ppb and 
$^{nat}$K is of order of few tens ppm.
The PMT contribution to the counting rate is further reduced by using 3'' diameter 10 cm long 
UV light guides (made of Suprasil B) which also act as optical windows, as mentioned above.
The highly radiopure light guides assure both suitable low radioactivity 
($< 1$ ppb of $^{238}$U, $< 2$ ppb of $^{232}$Th, and $< 1.3$ ppm of $^{nat}$K as measured by Ge detectors) 
and very suitable optical characteristics for the NaI(Tl) light emission, being transparent down to 
300 nm.

Finally, to further reduce the PMTs contribution to the background, in the DAMA/LIBRA apparatus
each PMT and related light guide are surrounded by a shaped low-radioactivity copper shield 
(see section \ref{p:shield} and Fig.~\ref{shield}b)).

\section{The electronic chain}   \label{sec_ele}

Each crystal of the DAMA/LIBRA apparatus is viewed by two PMTs with grounded cathode and supplied 
by 
positive high voltage.
The HV power supply for the PMTs is given by a CAEN multichannel voltage supply with 
voltage stability of 0.1\%.

\begin{figure}[!htb]
\centering
\includegraphics[width=0.9\textwidth]{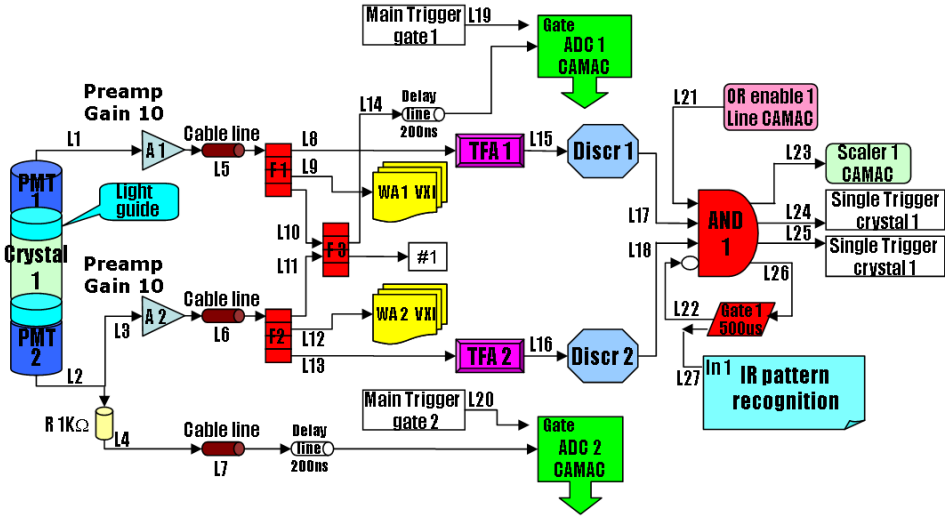}
\caption{The conceptual scheme of the electronic devices of a single detector. Devices processing
analog signals and those providing the trigger of the single detector are shown.}
\label{eletch1}
\end{figure}

Figure \ref{eletch1} shows the scheme of the electronic devices of a single detector. In particular,
the part of the chain regarding the analog signals and their processing are shown.
It is also reported the logic devices providing the trigger of the single detector.
The signals from each PMT are amplified by a preamplifier having 0-250 MHz 
bandwidth, a factor 10 gain and a voltage integral linearity $\pm 0.2\%$.

The signal from PMT 2 (see Fig. \ref{eletch1}) is divided in two branches by a resistive passive splitter: 
19/20 of the signal is sent to the preamplifier, while the remaining 1/20 
-- suitably delayed -- feeds the CAMAC charge ADC 2 ({\em QADC high energy}) unit. 
This latter allows the process of the high energy pulses, which instead saturate the 
remaining part of the electronics.
The preamplified signals of PMT 1 and PMT 2 are the inputs of linear Fan-in/Fan-out devices 
F1 and F2. They produce three copies of each signal:
i)   the first copies of PMT 1 and PMT 2 signals 
     are sent to two separate channels of a Waveform Analyser, which records
     the signals in a time window of 2048 ns. This is accomplished by using fast VXI Tektronix 
     four-channel TVS641A digitizers with a sampling frequency of 1 GSample/s and 250 MHz 
     bandwidth;
ii)  the second copies of PMT 1 and PMT 2 signals are added and sent to the input -- after a delay line -- 
     of the CAMAC charge ADC 1 ({\em QADC low energy}) and to a Spectroscopy Amplifier ({\it \#1}, see later);
iii) the last copies are used to generate the trigger of the crystal.
These latter ones are shaped by Timing Filter Amplifiers ($TFA1$ and $TFA2$) with an 
integration time of 50 ns; their outputs are discriminated by $DISCR1$ and $DISCR2$ with the thresholds
at the single photoelectron level.
The coincidence ({\it AND 1}) between the two logical NIM outputs provides the trigger of the considered detector.

In order to reject afterglows, Cherenkov pulses in the light guides and 
Bi-Po events, a 500$\mu$s veto occurs after each event selected by the coincidence 
({\it Gate 1} in Fig.~\ref{eletch1});
this veto introduces a negligible systematic error on the measured rate ($\lsim 10^{-4}$). 

\vspace{0.3cm}

A CAMAC Output Register ($OR$) provides the signal able to activate the coincidence module of each
detector; this feature can be used, for example, during the calibration runs to enable
the detector to be calibrated.

\begin{figure}[!htb]
\centering
\includegraphics[width=0.9\textwidth]{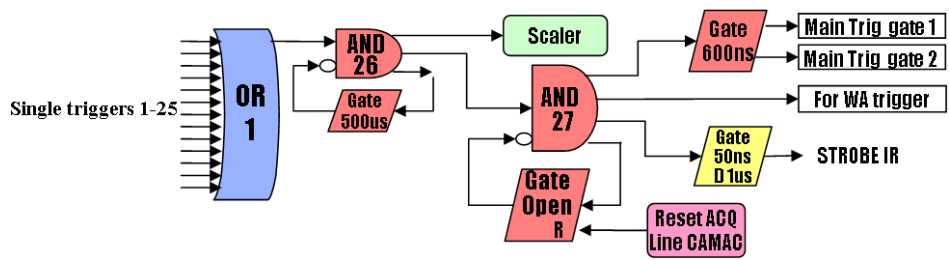}
\caption{The conceptual scheme of the electronic chain providing the main trigger of DAQ.}
\label{eletch2a}
\end{figure}

\begin{figure}[!htb]
\centering
\includegraphics[width=0.9\textwidth]{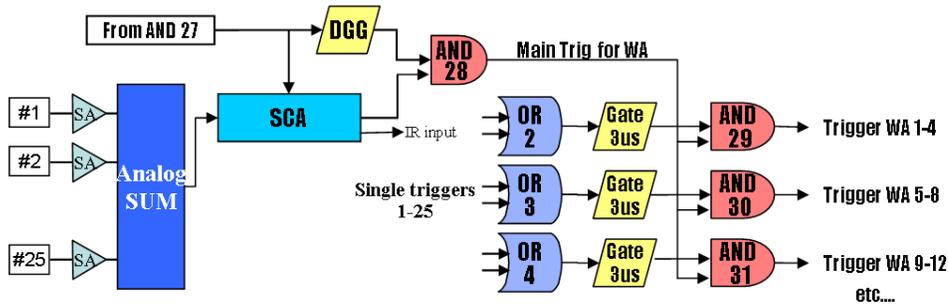}
\caption{The conceptual scheme of the logic system managing the individual triggers 
of the Waveform Analysers.}
\label{eletch2b}
\end{figure}

The outputs of the coincidence device (\mbox{{\it AND 1}}) provide: 
i)   the signal for a CAMAC scaler to count the number of triggers of each detector.
     The ratio of this number to the elapsed time gives the {\it hardware rate}, $R_{Hj}$ of the $j-$th detector; 
ii)  the line used in the main trigger electronic chain (see later and Fig.~\ref{eletch2a});
iii) the line used in the logic system managing the individual triggers of the Waveform Analysers
     (see later and Fig.~\ref{eletch2b});
iv)  the line giving the start to the {\it Gate 1} which -- in addition to the veto of the 
     coincidence \mbox{{\it AND 1}} -- feeds a CAMAC I/R Pattern Recognition ($IR$). The $IR$ allows
     the identification of the detector/detectors involved in the main trigger.

The electronic devices which generate the main trigger of the acquisition are shown in Fig.~\ref{eletch2a}.
The main trigger is provided by the logic $OR$ ({\it OR 1}) of all the crystals. The main trigger 
NIM logic pulses are counted by a Scaler.  
Devoted devices manage the trigger only when the acquisition is ready (see 
Fig.~\ref{eletch2a});
they issue the trigger of the DAQ.
The dead time of the acquisition is properly evaluated by using the information from the Scaler.
The live time is estimated straightforward.

When a DAQ trigger occurs, the following logic signals are issued to:
i)   the Gate Generator producing the 600 ns gates for the charge ADCs (QADC);
ii)  the Delay Gate Generator giving the strobe signal to the CAMAC I/R Pattern Recognition.
     This device also generates the LAM in the CAMAC system, and, therefore, 
     the interrupt to the CPU of the acquisition computer; 
iii) the logic system managing the individual triggers of the Waveform Analysers.

The logic system of the individual triggers of the Waveform Analysers
has been introduced in order to speed up the DAQ.
In particular, the trigger of the Waveform Analysers (see Fig.~\ref{eletch2b}) are issued 
only if the total energy deposited in the detectors is in an energy window suitably chosen.
For this purpose, each line feeds a Spectroscopy Amplifier ($SA$) whose gain is equalized.
The analog sum of the $SA$ signals of all the detectors, whose stability is monitored by a peak ADC
(not shown in the scheme), feeds a Single Channel Analyser ($SCA$) which allows the selection just of 
events in the chosen energy window.
A devoted electronic circuit issues the trigger of the Waveform Analysers\footnote{We remind that 
each
Waveform Analyser has four channels and, therefore, allocates the signals of two detectors.} 
allocating the detectors producing the main trigger. 
Summarizing, an individual trigger to a given Waveform Analyser is issued when the following three conditions 
are fulfilled: 
i)   at least one of its corresponding lines has fired; 
ii)  the DAQ trigger is present;
iii) the total energy of the events is in the chosen energy window.
Let us remind that for the events with energy outside the chosen energy window (e.g. high energy 
events) the QADC values are acquired anyhow.

\vspace{0.3cm}

\begin{figure}[!ht]
\centering
\includegraphics[width=0.75\textwidth] {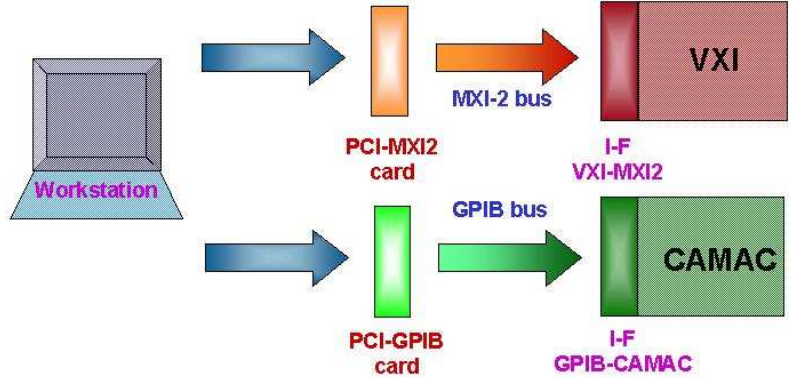}
\caption{Data acquisition system of DAMA/LIBRA.}
\label{fg:daq}
\end{figure}

The data acquisition system is made of a Workstation by Compaq with Linux Operating
System, interfaced with the hardware system through MXI-2 and GPIB buses (see Fig. \ref{fg:daq}).
The GPIB bus allows the communication with the CAMAC crate housing the QADCs, the scalers and 
the I/O registers, while the MXI-2 bus allows to communicate with the three VXI mainframes, 
where the Waveform Analysers are installed.

The information collected by DAQ, when a DAQ trigger occurs, are: i) the time of occurrence of the event;
ii) the pattern of the detectors firing; iii) for each fired detector the QADC channels and 
the pulse profiles of the two PMT's of the detector as recorded by the Waveform Analyser,
when the conditions given above are fulfilled.
In particular, the areas of the pulse profiles ({\it TD channel}) are proportional to the 
energy released in the detector (see Sect. \ref{sec_gf}). Other variables can be constructed by
the pulse profiles and they can be used for noise rejection (see Sect. \ref{sec_gf}).
For high energy events the energy can be estimated in terms of QADC channels (see Sect. \ref{sec_gf}).

Moreover, let us remind that the DAQ also records -- together with the production data --
the information coming from the monitoring system, as mentioned above (also see Fig. \ref{fg:proc}).

In near future the DAMA/LIBRA DAQ will be improved and new Waveform Analysers will be introduced.
In particular, in the new DAQ a HP Workstation with Intel processor (3.4 GHz) and Red-Hat 
Linux Operating System will be interfaced with the hardware system through MXI-3 and GPIB 
buses. The new MXI-3 bus will communicate with the new Waveform Analysers through a 
CompactPCI mainframe.
The new Waveform Analysers will be Quad-Channel CompactPCI Digitizer Acqiris DC270 with 250 
MHz bandwidth and a sampling frequency of 1 GSample/s simultaneously on all 4 channels.
The data will be transferred on fiber optics from the Waveform Analysers to PC (electrically decoupled).

\section{General features}      \label{sec_gf}

\vspace{0.1cm} 
\noindent {\bf Uniformity of the light collection}
\vspace{0.1cm}

The absence of dead spaces in the light collection has been carefully investigated by performing 
suitable calibrations in different positions of the detector; in fact,
by irradiating the whole detector with high-energy 
$\gamma$ sources (e.g., $^{137}$Cs) from different positions, no significant variations of the peak 
position and energy resolution have been observed.
Moreover, the $\alpha$ peaks at high energy (see Fig. \ref{timamp})
and their energy resolutions are well compatible with those expected for $\gamma$ calibration (see above)
and also support the uniformity of the light collection within 0.5\%\footnote{For example,
from Fig. \ref{timamp} $a)$ the energy resolution is $\sigma = (75 \pm 3)$ keV, while the expected one
from the fit of Fig. \ref{fg:cal_he} is 72 keV. Hence, with the help of a simulation of the light 
collection, an upper limit of about 0.5\% can be set on the disuniformity of the light collection.}.

\vspace{0.1cm} 
\noindent {\bf Photoelectrons/keV}
\vspace{0.1cm}

The number of photoelectrons/keV has been derived for each detector 
from the information collected by the Waveform Analyser
over its large time window (2048 ns). 

\begin{figure}[!ht]
\centering
\vspace{-0.4cm}
\includegraphics[width=6.cm] {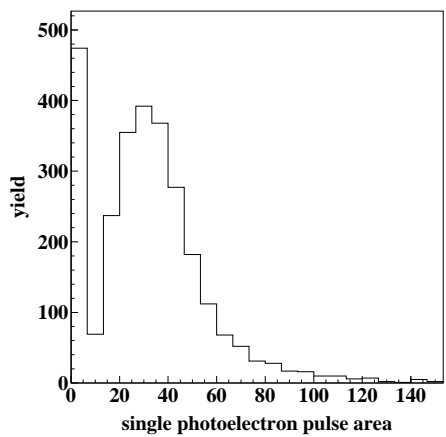}
\vspace{-0.4cm}
\caption{Typical experimental distribution of the area of the single photoelectron's pulses for a
DAMA/LIBRA detector, as recorded by the Waveform Analyser over 2048 ns time window. 
From this picture 6.7 photoelectrons/keV have been inferred for this detector.}
\vspace{-0.2cm}
\label{fg:photoel}
\end{figure}

In fact, a clean sample of photoelectrons can be extracted from the end part of this time window; 
there the scintillation pulses are completely ended, while afterglow single photoelectron
signals can be present.
As an example, Fig. \ref{fg:photoel} shows a typical experimental distribution of the area of the 
single photoelectron's pulses for a DAMA/LIBRA detector.
The relative peak value can be compared with the peak position of the distribution of the areas 
of the pulses corresponding to a full energy deposition from the 59.5 keV of the $^{241}$Am source. 
This procedure permits to obtain the number of photoelectrons/keV searched for;
they ranges from 5.5 to 7.5 photoelectrons/keV depending on the detector.

\vspace{0.8cm} 
\noindent {\bf Calibrations at low and high energy}
 \vspace{0.1cm}

\begin{figure}[!ht]
\centering
\includegraphics[width=5.cm] {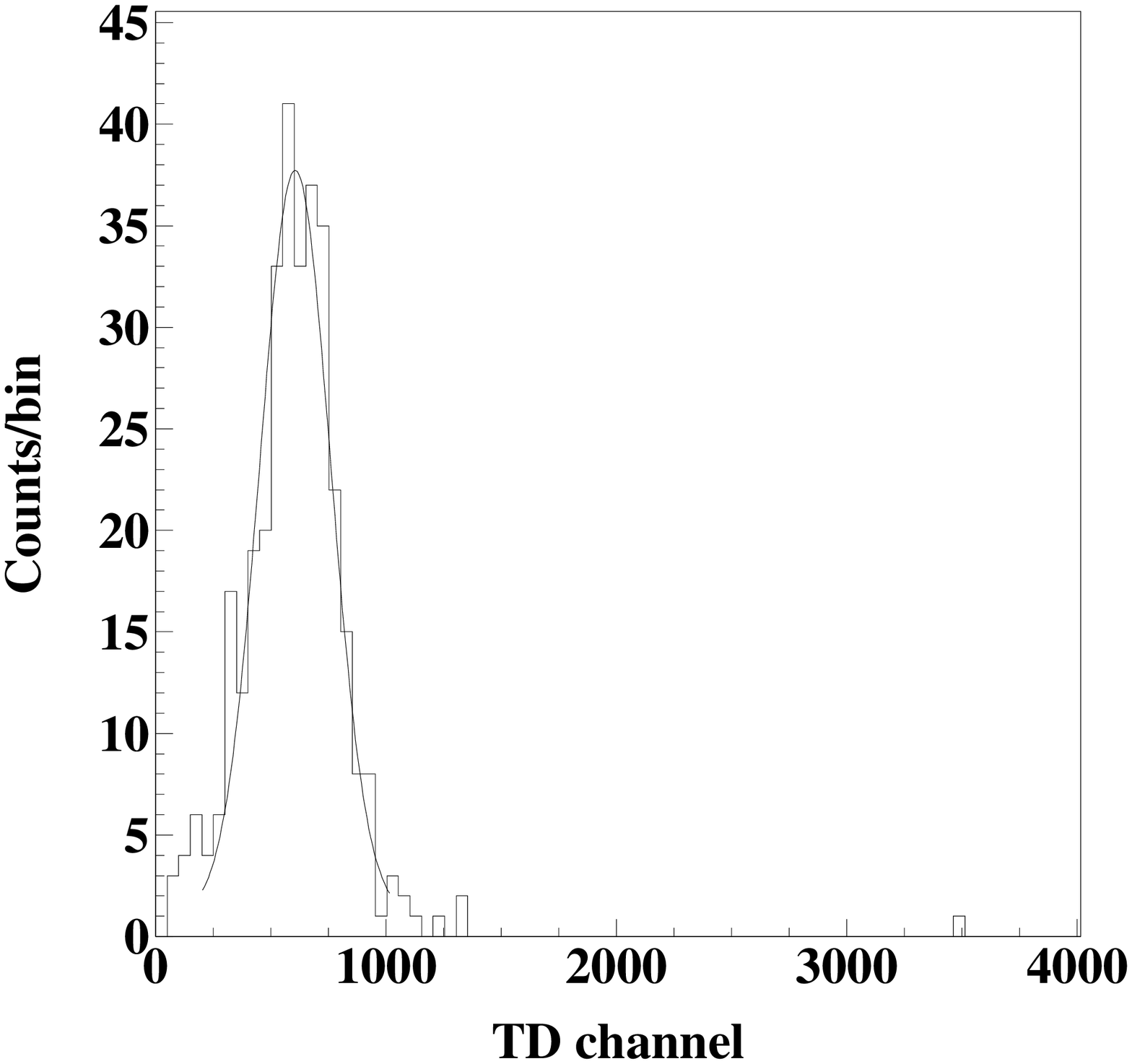}
\includegraphics[width=5.cm] {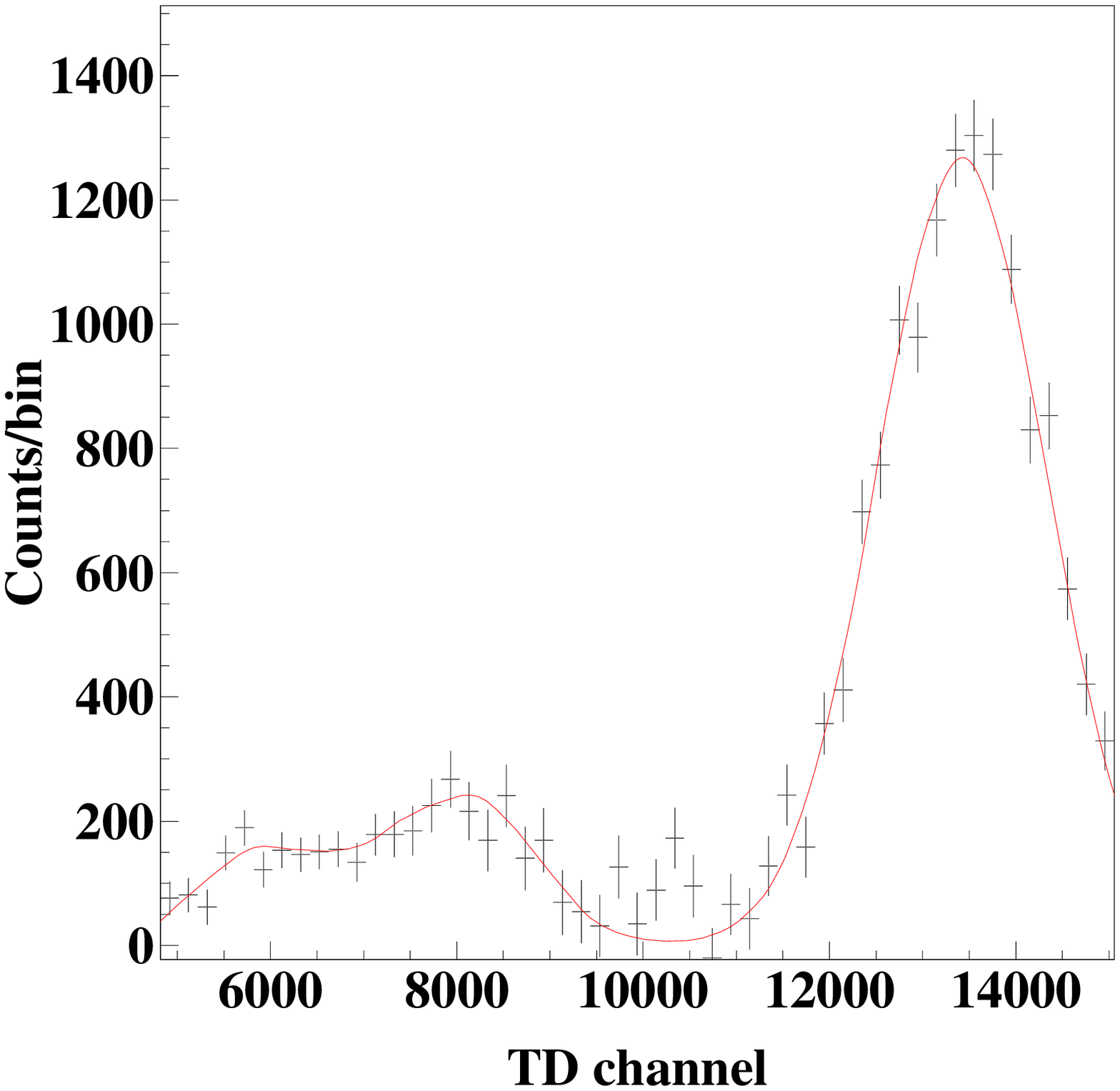}
\includegraphics[width=5.cm] {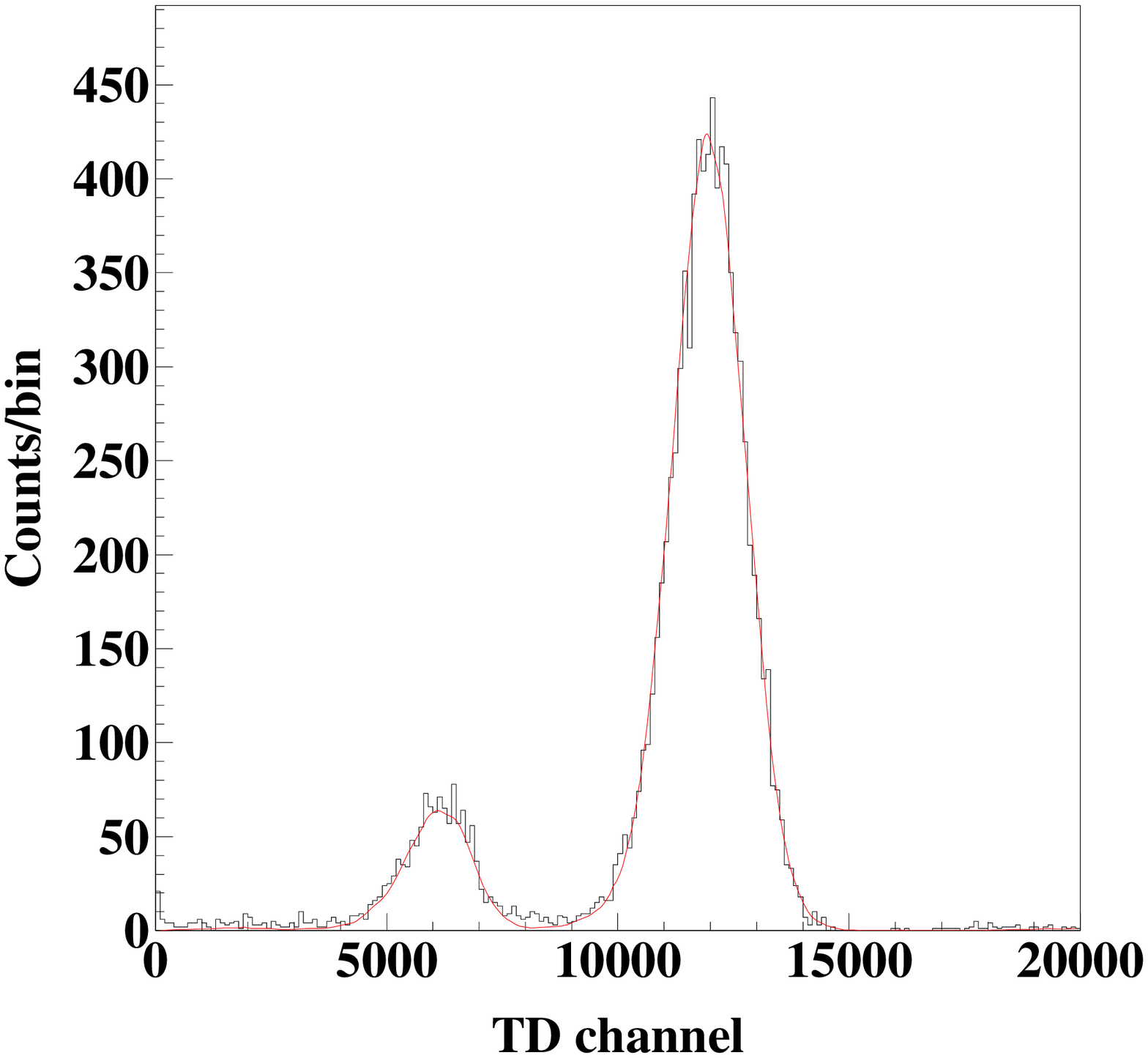}
\includegraphics[width=5.cm] {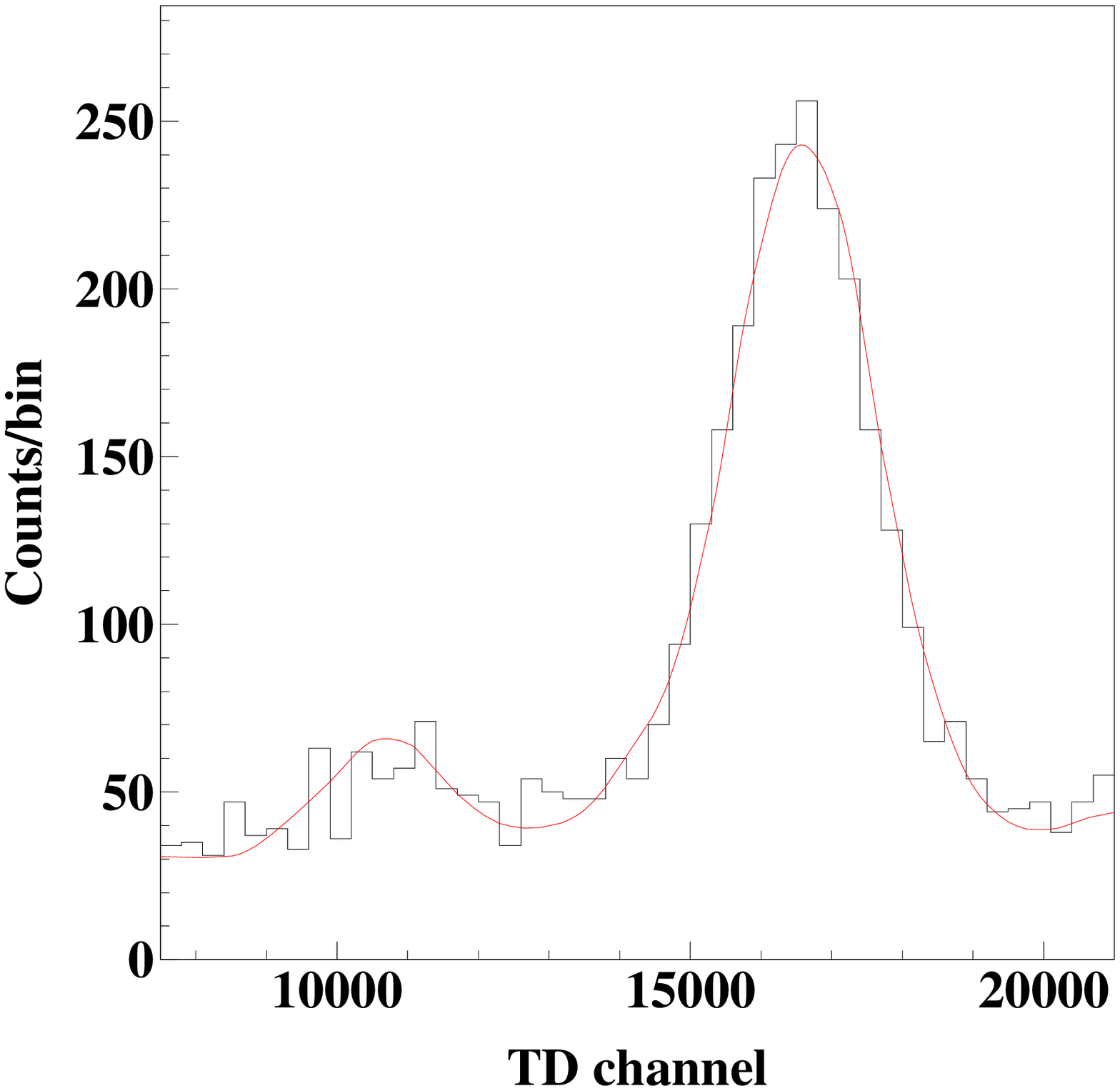}
\vspace{-0.2cm}
\caption{Examples of energy (pulse area) distributions for: 
i) X-rays/Auger electrons (3.2 keV) due to internal $^{40}$K ({\it top left}); 
ii) gammas and/or X-rays and/or Auger electrons -- mainly two structures in the energy 
spectrum at 40.4 keV (composite peak) and at 67.3 keV -- due to internal $^{125}$I ({\it top right}); 
iii) gammas from $^{241}$Am external source ({\it bottom left}); 
iv) gammas of 81 keV from $^{133}$Ba external source ({\it bottom right}). 
The lines superimposed to the experimental data 
have been obtained by MonteCarlo simulations, obviously taking into account
the geometry and all the involved materials.}
\label{fg:calib}
\end{figure}

The low (up to $\sim 100 $ keV) and the high (above $\sim 100$ keV) energy regions
have been studied by using various external gamma sources and internal X-rays or gamma's. 
In particular, as shown in Fig.~\ref{fg:calib}, gammas from $^{241}$Am and from $^{133}$Ba
external sources provide 30.4 keV (composite), 59.5 keV and 81.0 keV peaks;
moreover, other calibration points can be obtained by: 
i)   the internal X-rays at 3.2 keV (see Fig. \ref{k40_8} and Fig. \ref{fg:calib});
ii)  the gammas and/or X-rays and/or Auger electrons due to internal $^{125}$I 
     (see Fig. \ref{fig70} $a)$ and Fig. \ref{fg:calib}), giving two structures in the energy 
     spectrum at 40.4 keV (composite peak) and at 67.3 keV;
iii) the gammas at 39.6 keV plus the $\beta$ spectrum due to cosmogenic $^{129}$I (see Fig. \ref{fg:129i}).

In running conditions routine calibrations with $^{241}$Am sources are regularly carried out.

\begin{figure}[!hb]
\centering
\includegraphics[width=6.cm] {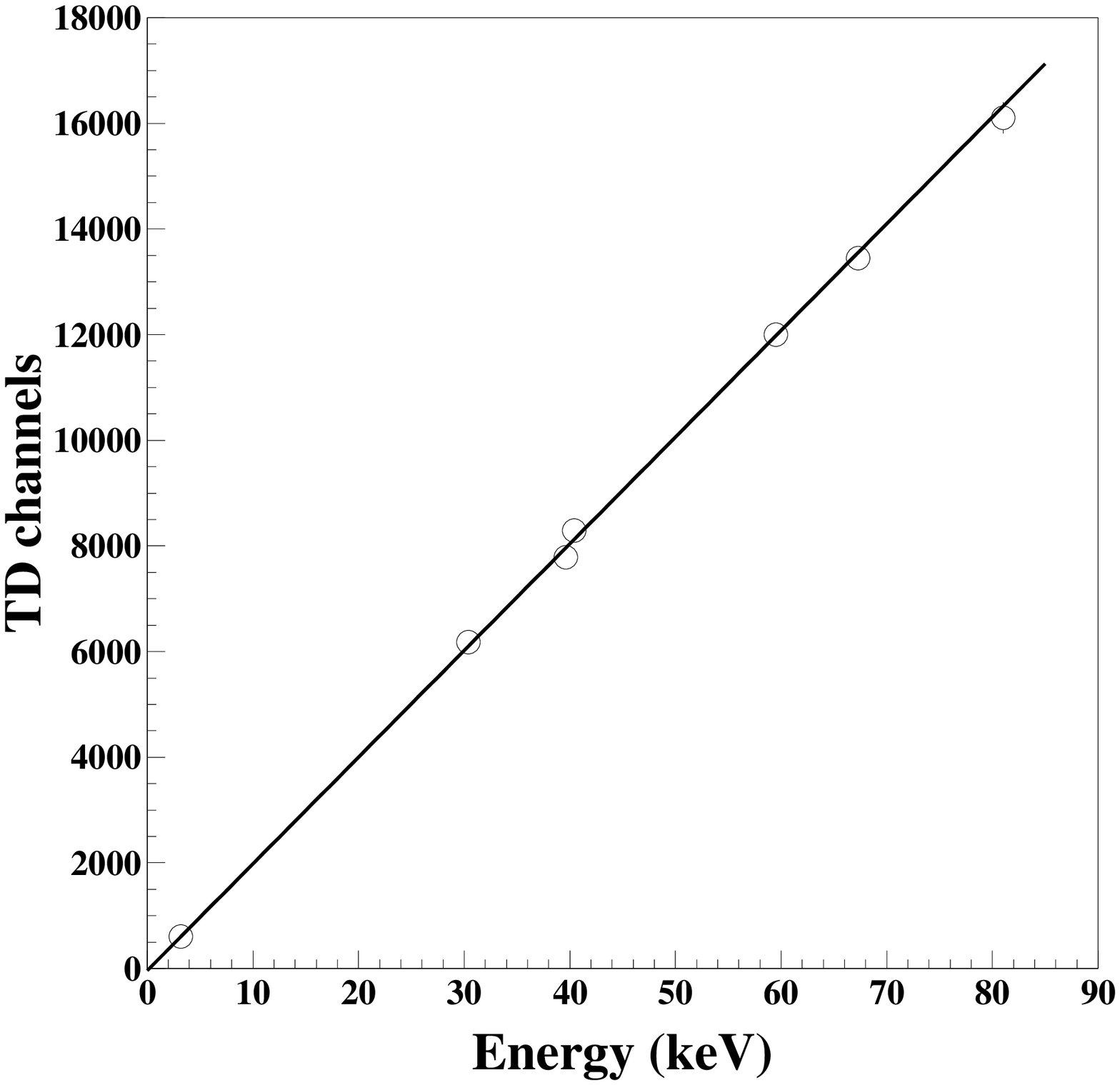}
\includegraphics[width=6.cm] {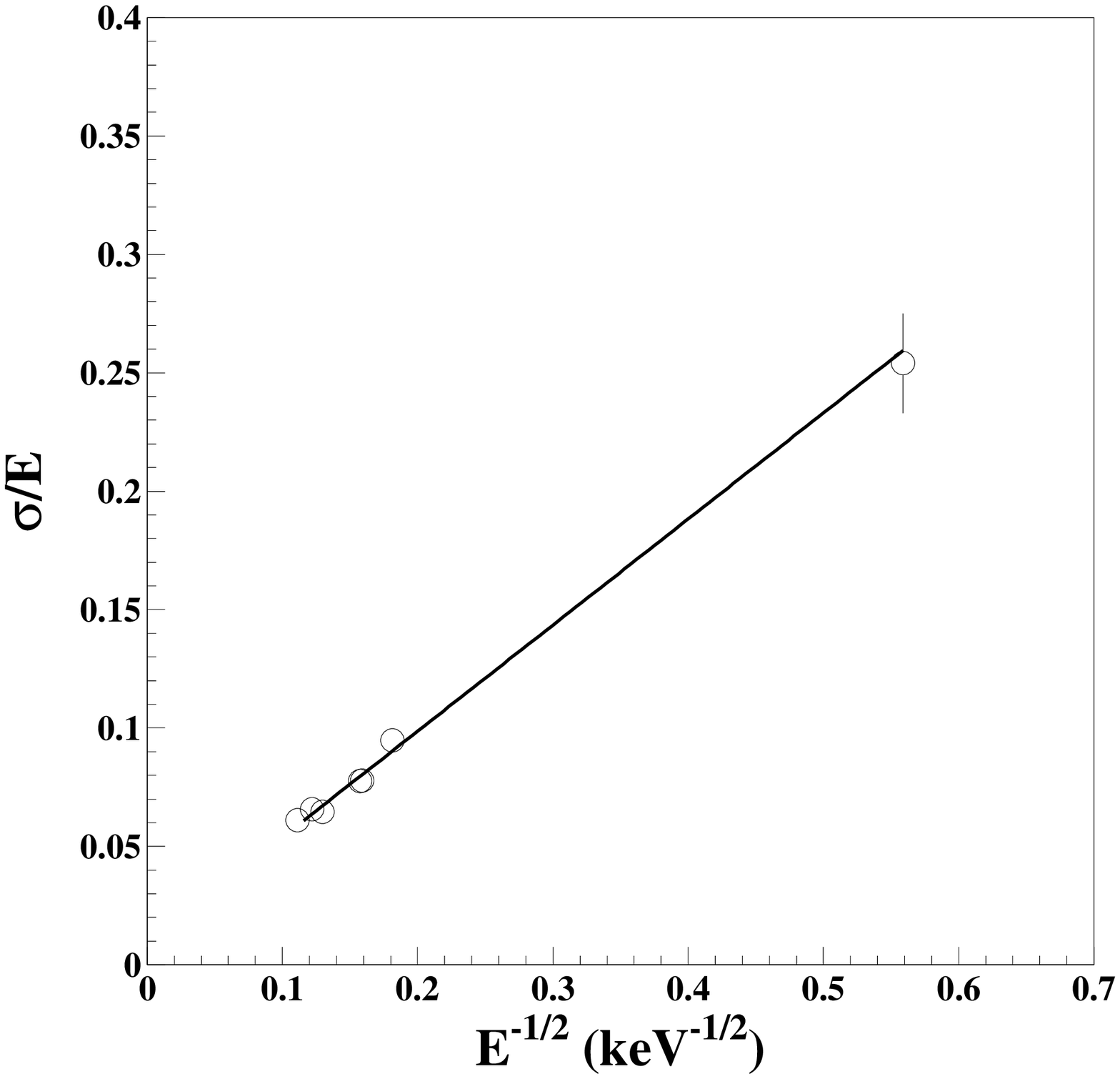}
\vspace{-0.2cm}
\caption{Linearity of the pulse area as a function of the energy 
and energy resolution ($\sigma/E$) as a function of $E^{-1/2}$
for low energy calibrations (internal and external radiation).}
\vspace{-0.1cm}
\label{fg:linres}
\end{figure}

The linearity of the pulse area as a function of the energy of the calibration points 
and the energy resolution in the low energy region are depicted in Fig. \ref{fg:linres}.
The energy resolution behavior for the low energy scale 
has been fitted by a straight line:
$\sigma_{LE}/E = \frac{\alpha_{LE}}{\sqrt{E(keV)}} + \beta_{LE}$; the best fit values are
$\alpha_{LE} = (0.448\pm0.035)$ and $\beta_{LE} = (9.1\pm5.1)\times10^{-3}$.


\begin{figure}[!t]
\centering
\includegraphics[width=5.cm] {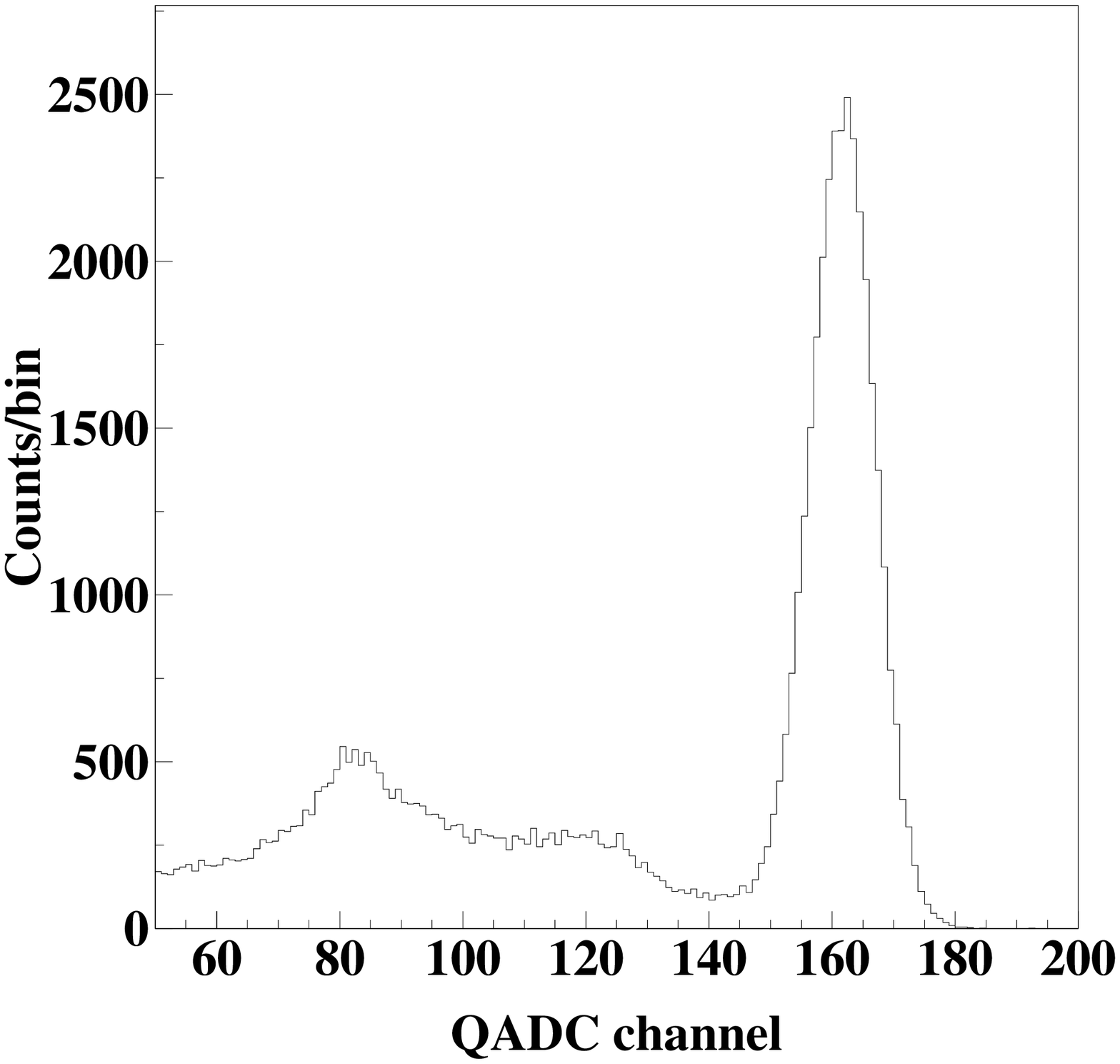}
\includegraphics[width=5.cm] {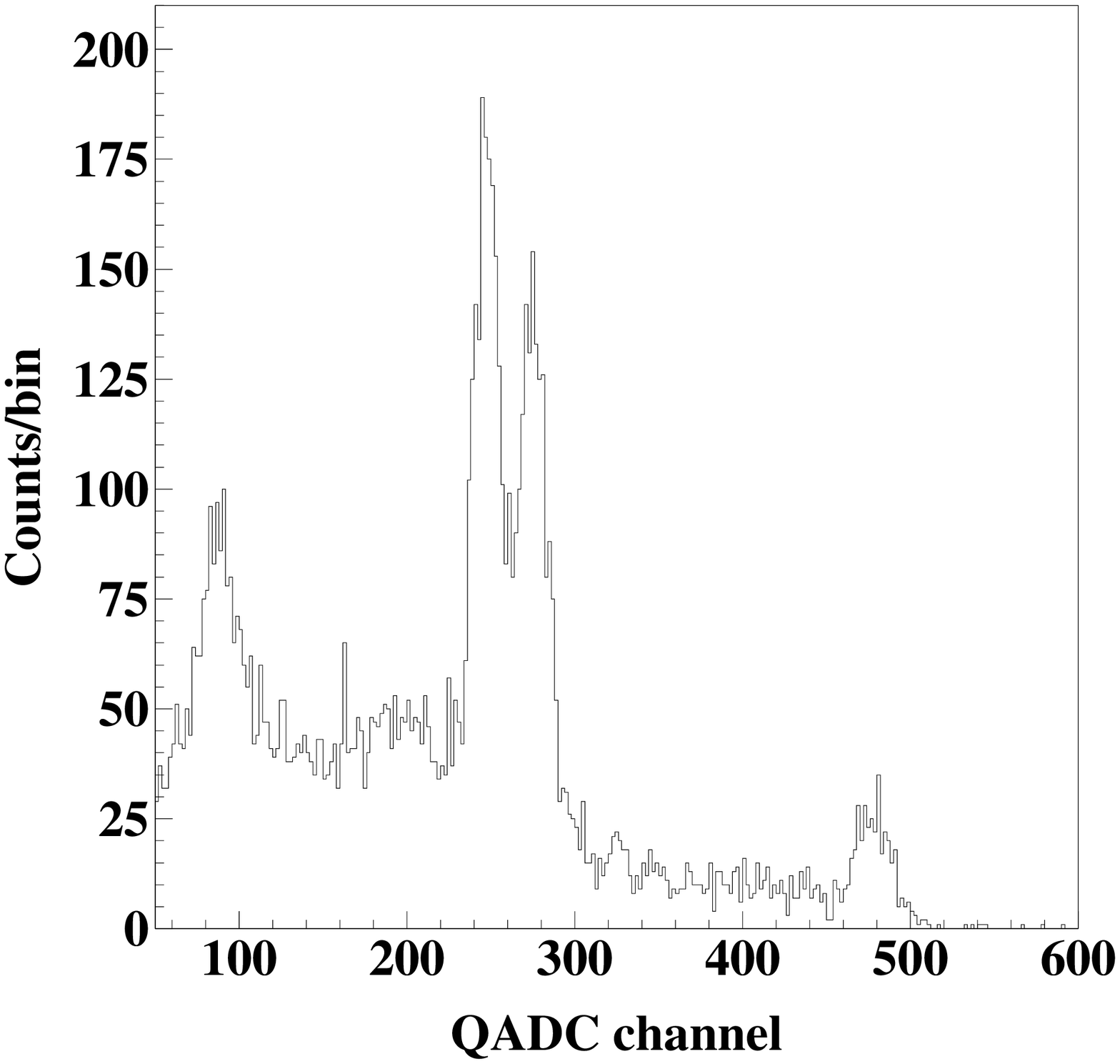}
\includegraphics[width=5.cm] {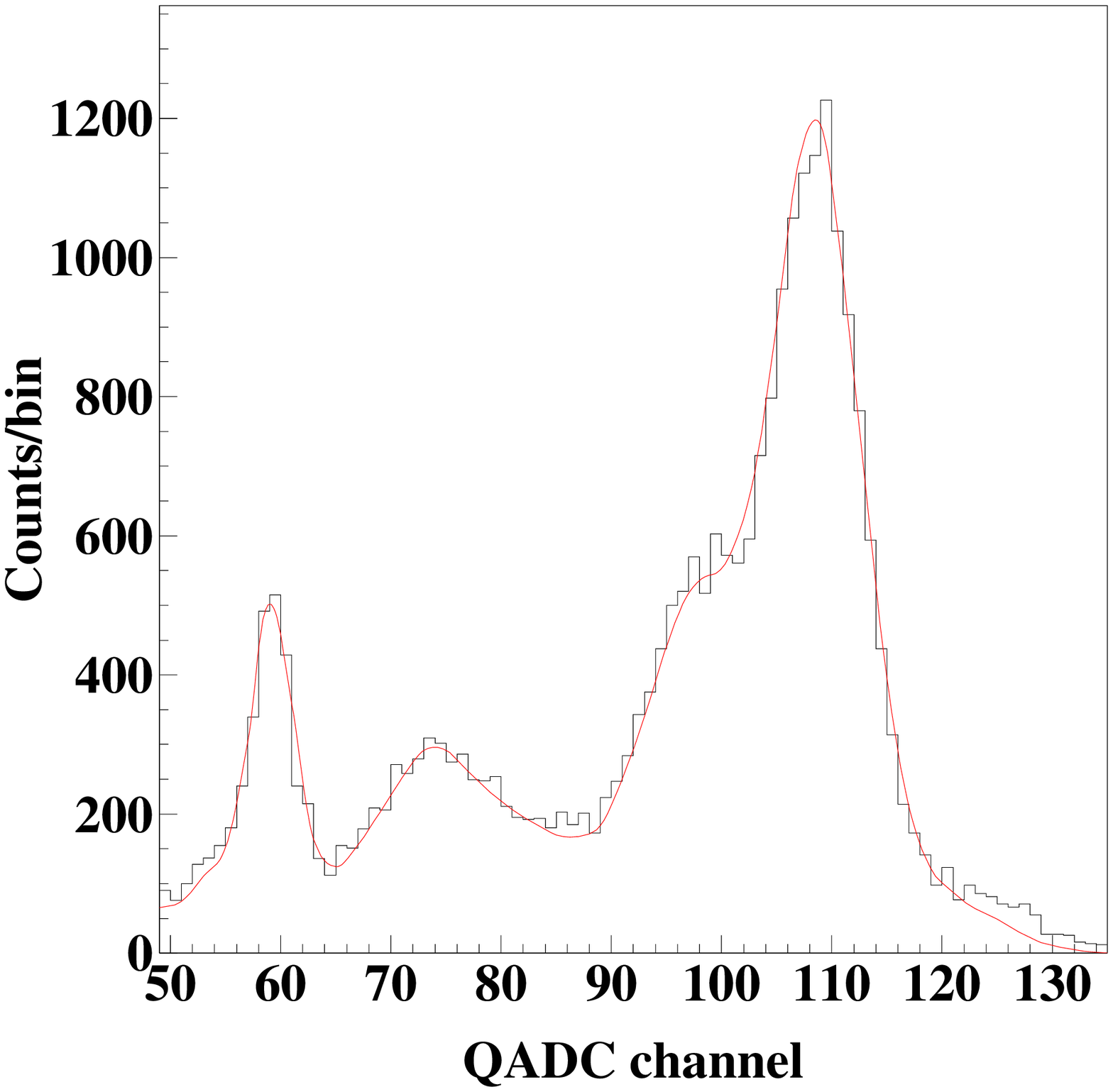}
\includegraphics[width=5.cm] {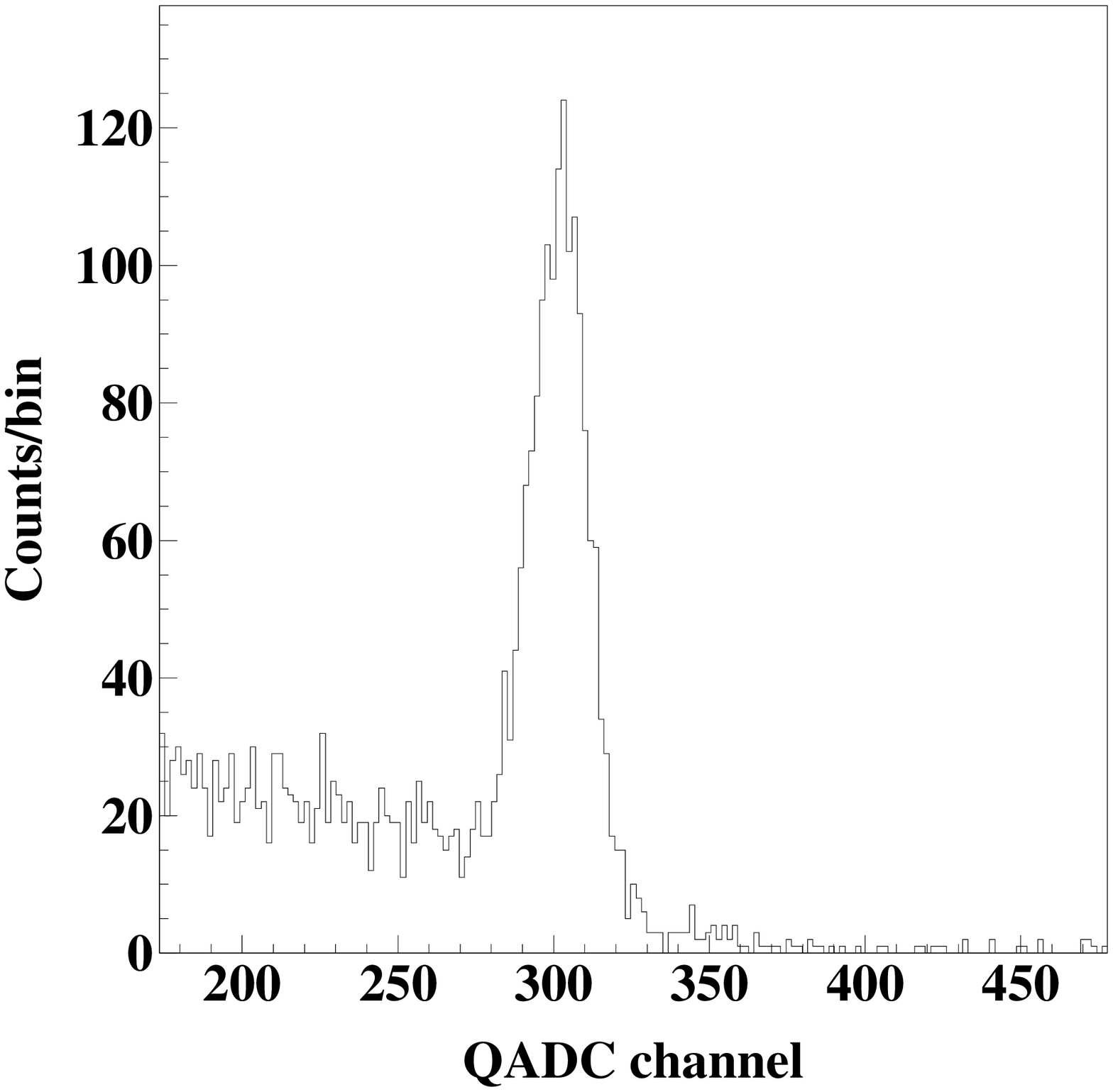}
\vspace{-0.2cm}
\caption{Energy distributions for the external sources: $^{137}$Cs, 662 keV $\gamma$ rays ({\it top left}); 
$^{60}$Co, 1173 keV and 1332 keV $\gamma$ rays and their sum: 2505 keV ({\it top right});
$^{133}$Ba, mainly 81 keV and 356 keV $\gamma$ rays ({\it bottom left}).
Energy distribution for $\gamma$ rays of 1461 keV ({\it bottom right}) 
due to $^{40}$K decays, tagged by the 3.2 keV X-rays in an adjacent detector 
(see also Fig. \ref{k40_8}).
The line superimposed to the experimental data of the $^{133}$Ba spectrum
has been obtained by MonteCarlo simulations.
Here in abscissa the QADC channel is reported.}
\label{fg:spe_he}
\end{figure}

\begin{figure}[!ht]
\centering
\vspace{-0.5cm}
\includegraphics[width=6.cm] {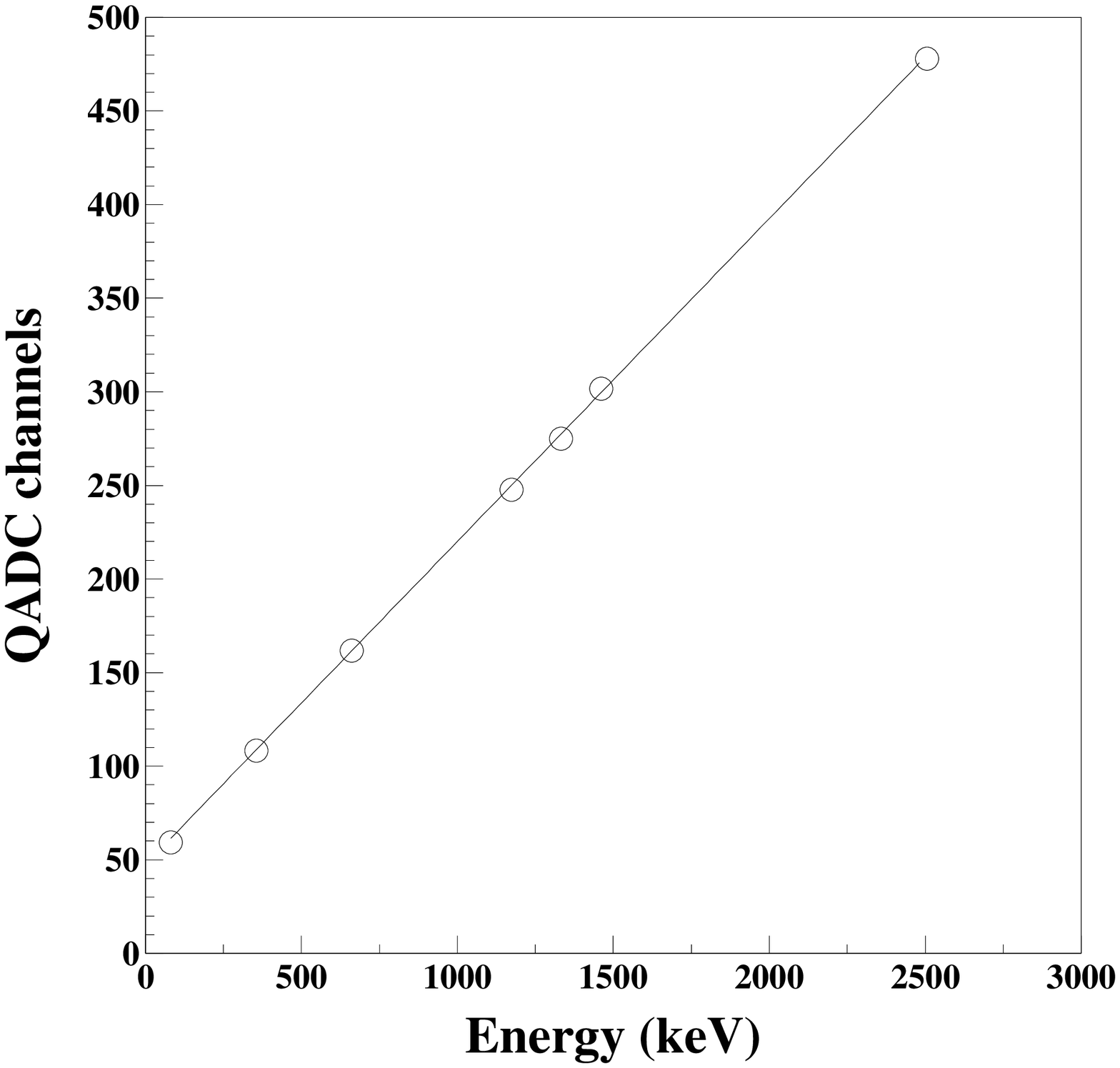}
\includegraphics[width=6.cm] {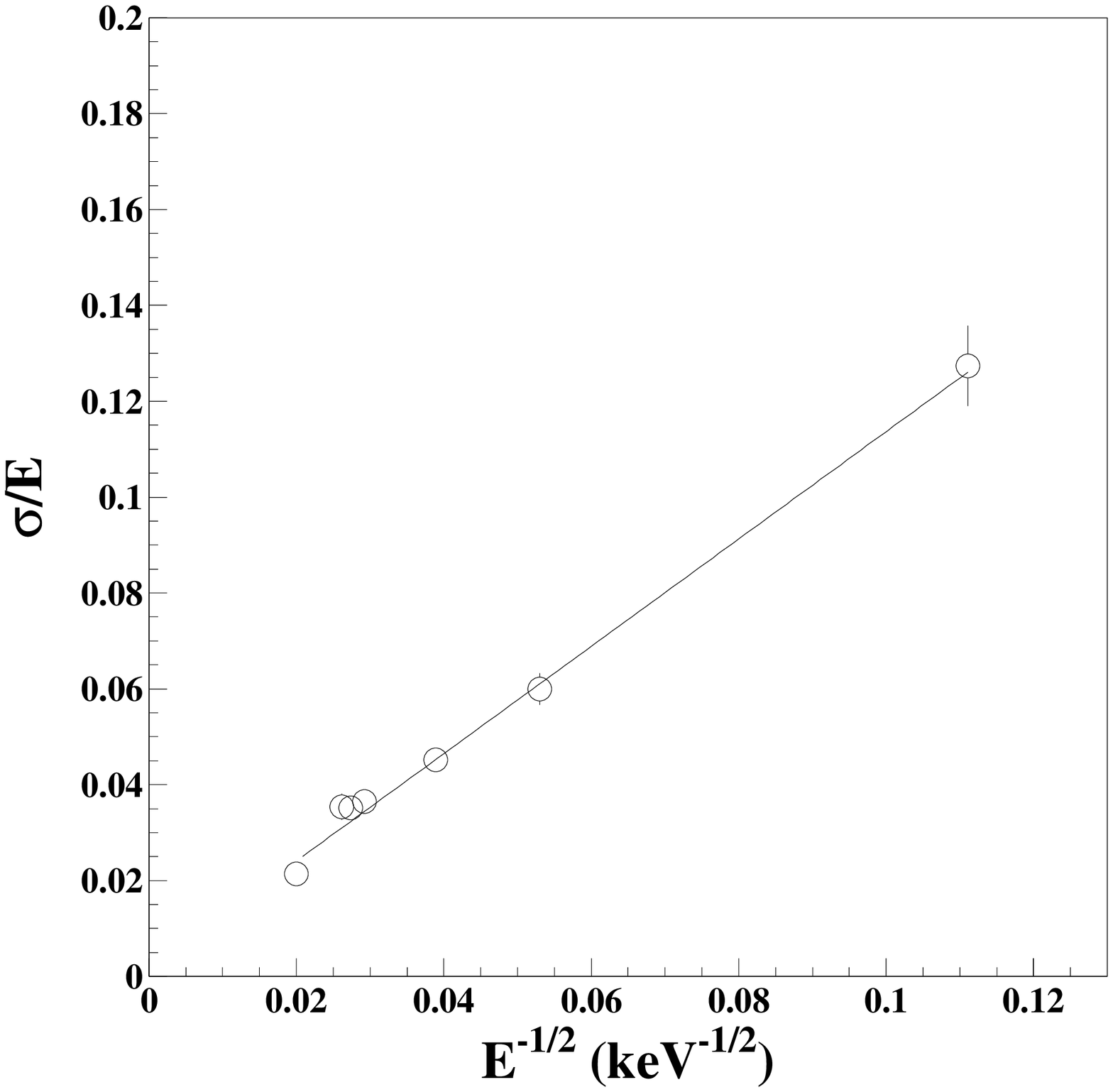}
\vspace{-0.8cm}
\caption{Linearity of the QADC as a function of the energy 
and energy resolution ($\sigma/E$) as a function of $E^{-1/2}$
in the high energy region. We remind that the signals 
-- unlike low energy events -- for high energy events are taken only from one PMT.}
\label{fg:cal_he}
\end{figure}

We remind here that, although the optimization is made for the lowest energy region,
the data are taken on the full energy scale up to the MeV region by means QADC's.
Thus, the high energy scale of the QADC has been calibrated mainly by using external sources
of $\gamma$ rays, such as: $^{137}$Cs (662 keV), $^{60}$Co (1173 keV and 1332 keV and 
their sum: 2505 keV) and $^{133}$Ba (mainly 81 keV and 356 keV); examples are reported in
Fig. \ref{fg:spe_he}.
Moreover, $\gamma$ rays of 1461 keV due to $^{40}$K decays, tagged by the 3.2 keV X-rays 
in an adjacent detector (see also Fig. \ref{k40_8}), have been used.

The linearity of the QADC as a function of the energy 
and the energy resolution in the high energy region are reported in Fig. \ref{fg:cal_he}.
The energy resolution behavior for the high energy scale
has been fitted by a straight line:
$\sigma_{HE}/E = \frac{\alpha_{HE}}{\sqrt{E(keV)}} + \beta_{HE}$; the best fit values are
$\alpha_{HE} = (1.12\pm0.06)$ and $\beta_{HE} = (17\pm23)\times10^{-4}$.
The value $\alpha_{HE}$ is larger than the corresponding value at low energy since -- as discussed 
previously -- the signals for high energy events are taken only from one PMT. 

\begin{figure}[!ht]
\centering
\includegraphics[width=6.cm] {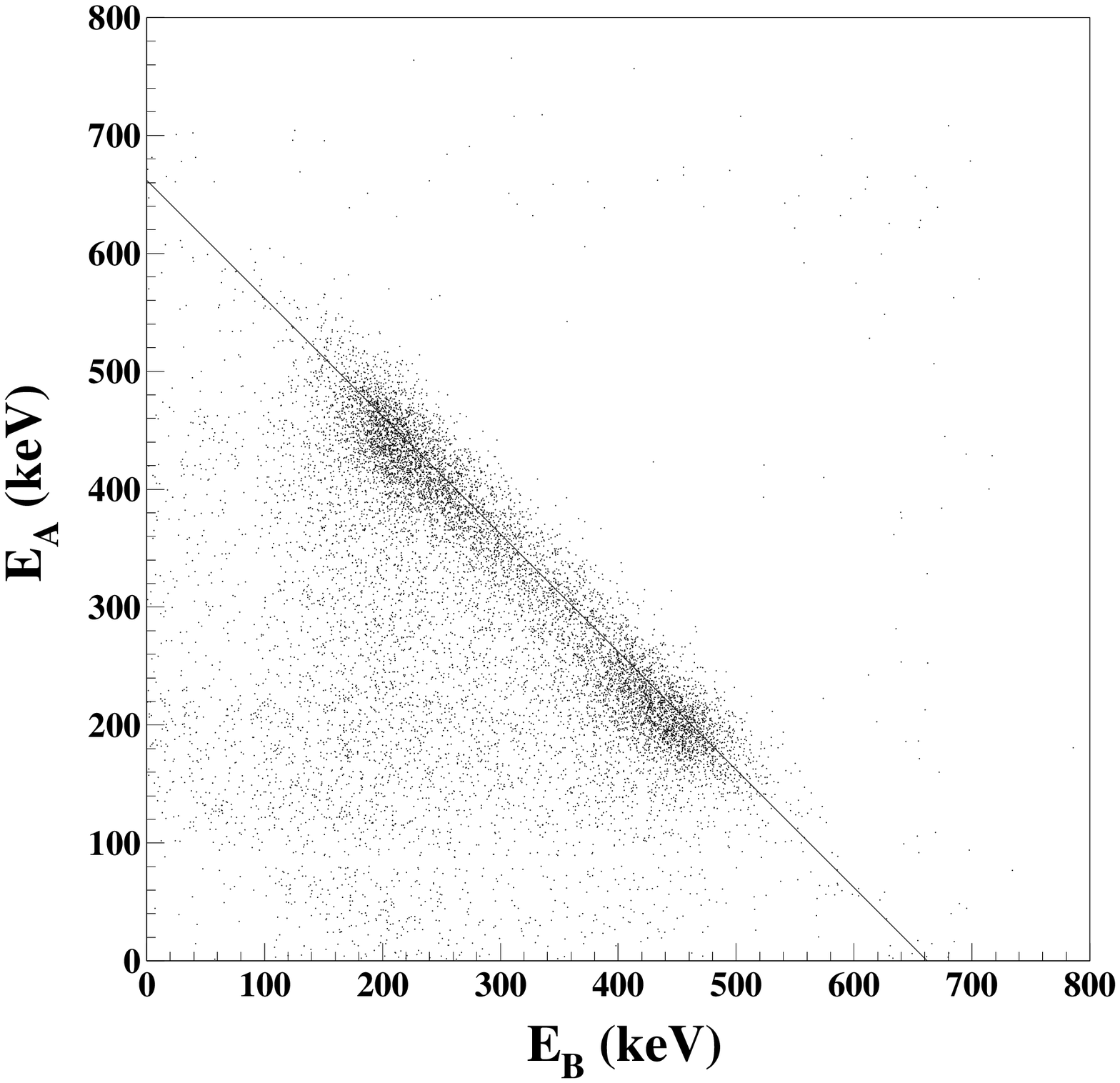}
\includegraphics[width=6.cm] {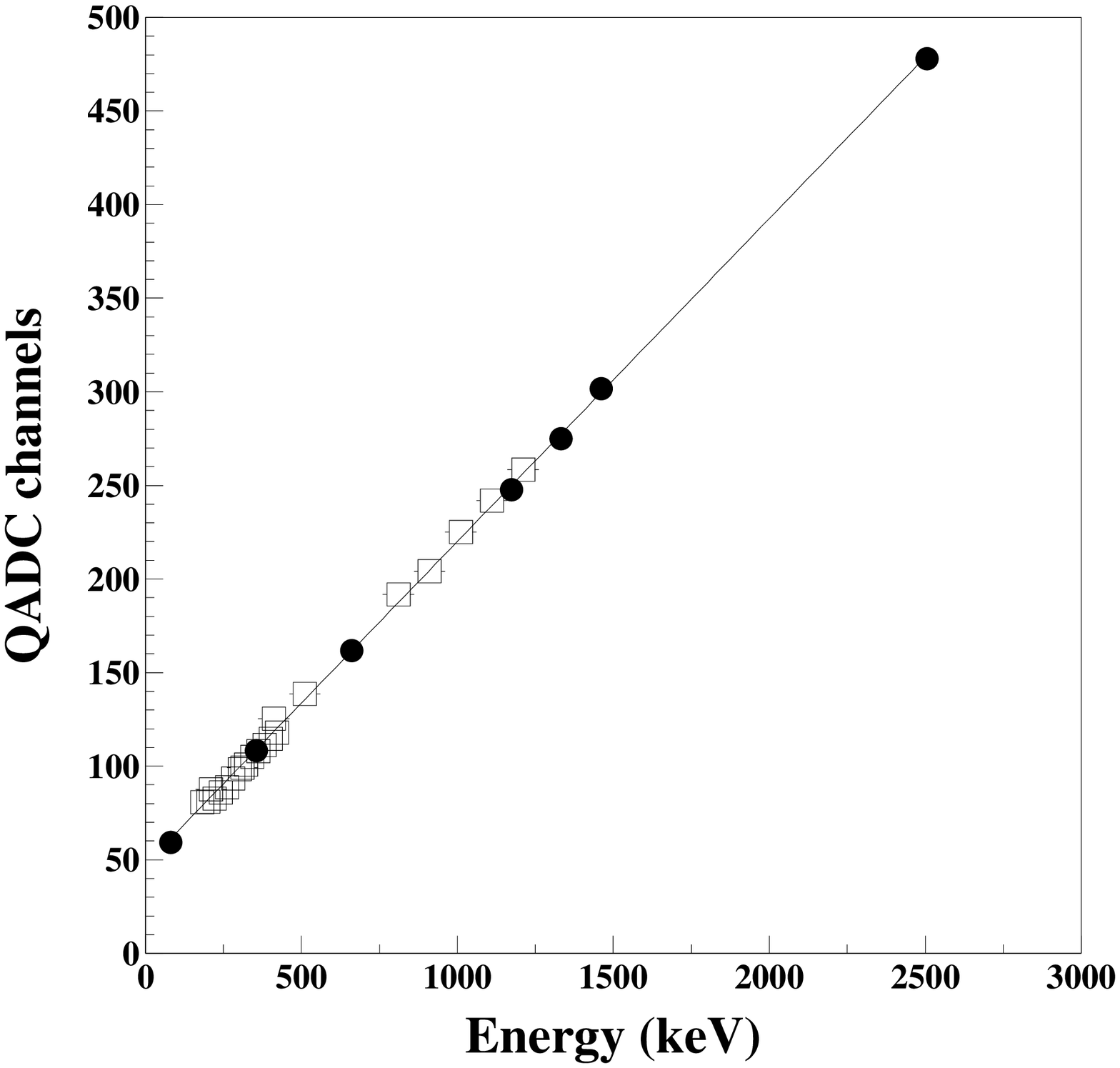}
\caption{{\it Left:} Example of a scatter plot of the energies, $E_A$ and $E_B$, of two detectors
$A$ and $B$
when using an external $^{137}$Cs source placed in between. The solid line
represents the case $E_A+E_B=662$ keV, energy of the $\gamma$ rays.
The data points around this line correspond to events where the $\gamma$ has 
a Compton back-scattering in one detector and the scattered $\gamma$ is completely
absorbed in the other one. 
{\it Right:} Linearity of the QADC as a function of the energy: data points from 
Fig. \ref{fg:cal_he} (filled points) and from tagged $\gamma$'s by double coincidences
as described in the text (squared points).}
\label{fg:lin}
\end{figure}

Another method to study the linearity is to use events of double coincidences when 
two detectors are irradiated by $\gamma$'s of known energy, $E_\gamma$. 
As an example, Fig. \ref{fg:lin}-{\it left}
shows the scatter plot of the energies, $E_A$ and $E_B$, of the two detectors $A$ and $B$
when using an external $^{137}$Cs source placed in between. The solid line
represents the case $E_A+E_B=E_\gamma$;
the data points around this line correspond to events where the 662 keV $\gamma$ has 
a Compton back-scattering in one detector and the scattered $\gamma$ is completely
absorbed in the other one. Fixing a slice -- for example at a fixed $E_B$ value -- 
it is possible to extract the peak position on the $E_A$ variable.
Applying this procedure to the high energy calibration data from various sources
other points can be added in the linearity plot, as done in Fig. \ref{fg:lin}-{\it right}.

\vspace{0.1cm} 
\noindent {\bf Procedures for noise rejection near the energy threshold}
\vspace{0.1cm}

The only data treatment which is performed on the raw
data is to eliminate obvious noise events (whose number
sharply decreases when increasing the number of available
photoelectrons) near the energy threshold. 
In particular, the DAMA/LIBRA detectors are seen by two PMTs working in coincidence and
this already strongly reduces the noise near the energy threshold for the {\it single-hit} events
(of interest for Dark Matter particles detection), while obviously noise is practically 
absent in {\it multiple-hit} events (see e.g. the double coincidence case of Fig. \ref{k40_8}),
since the probability to have random coincidences is negligible ($<3\times 10^{-6}$).

Moreover, the NaI(Tl) is a suitable detector for an effective noise rejection, having no contribution from 
microphonic noise (as it is instead the case e.g. of ionizing and bolometer detectors) and being the physical 
pulses well distinguishable from the noise pulses, essentially PMT noise.
This effectiveness is higher when a high number of photoelectrons/keV is available
and when PMTs with suitable performances have been built, as in the present case.

\begin{figure}[!ht]
\centering
\includegraphics[width=8.cm] {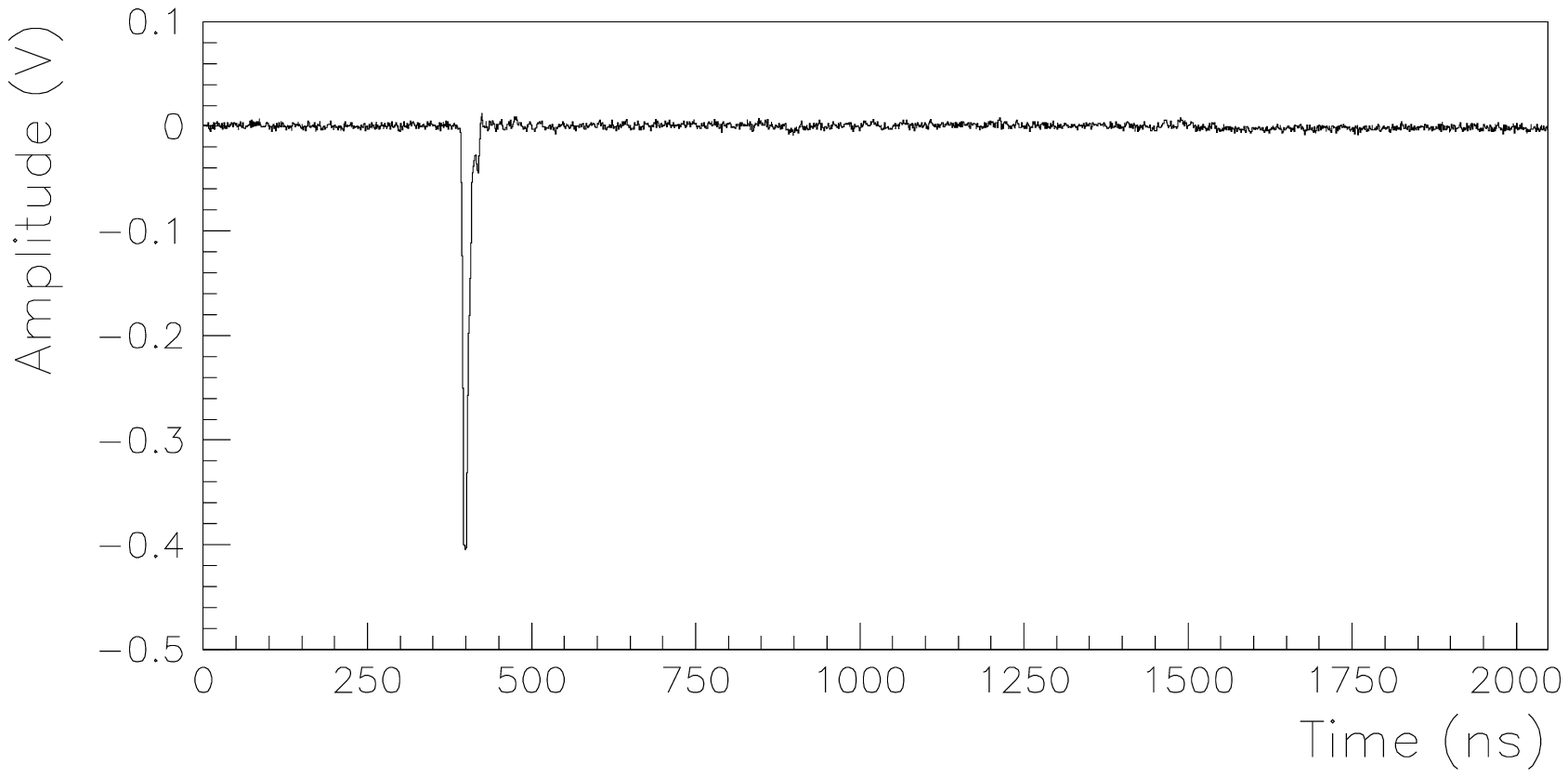}
\includegraphics[width=8.cm] {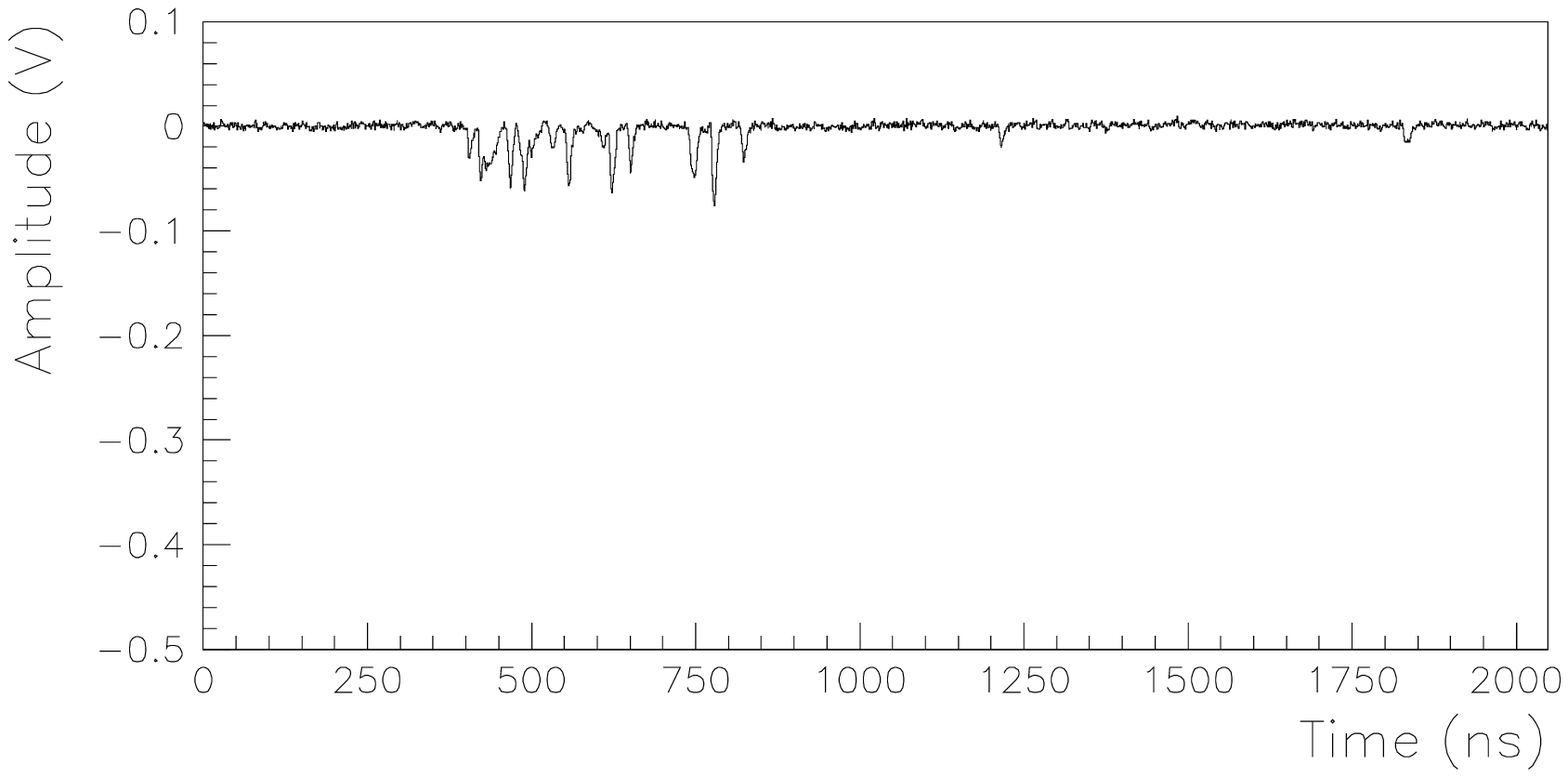}
\caption{Typical pulse profiles of PMT noise ($top$) and of scintillation event ($bottom$)
with the same area, just above the energy threshold of 2 keV.}
\label{fg:pulses}
\end{figure}

\begin{figure}[!ht]
\centering
\vspace{-0.8cm}
\includegraphics[width=6.cm] {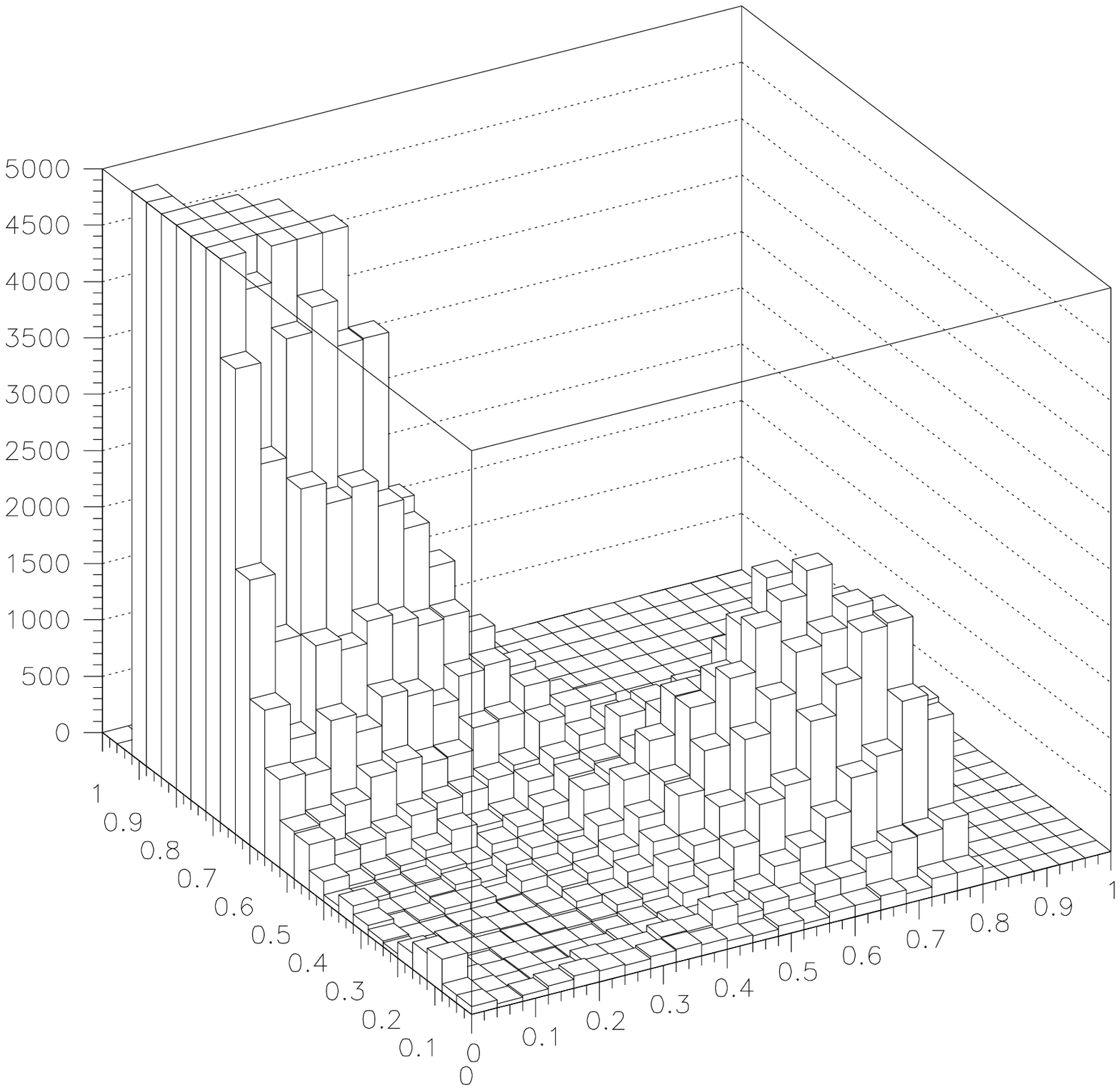}
\includegraphics[width=6.cm] {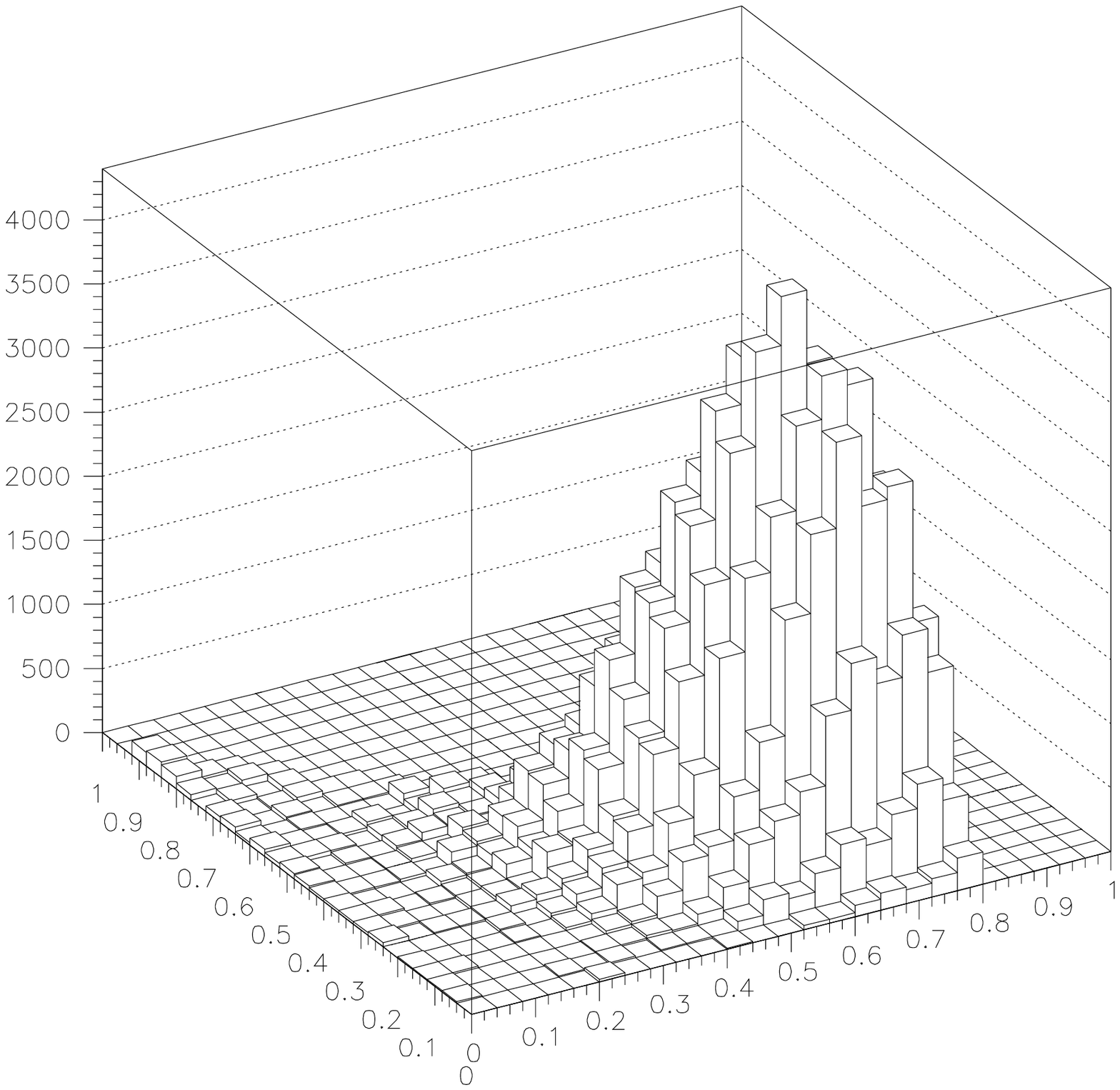}
\includegraphics[width=6.cm] {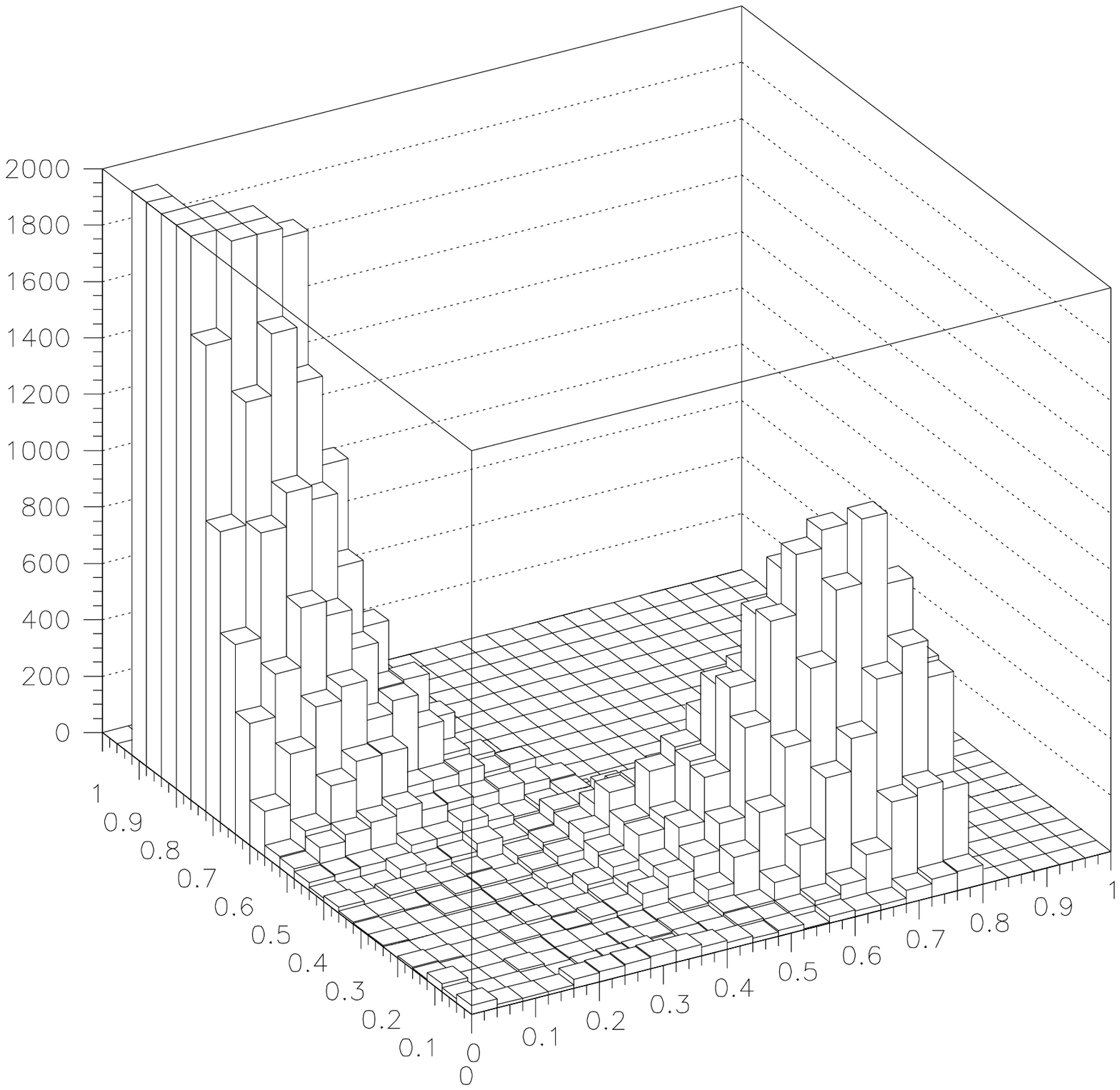}
\includegraphics[width=6.cm] {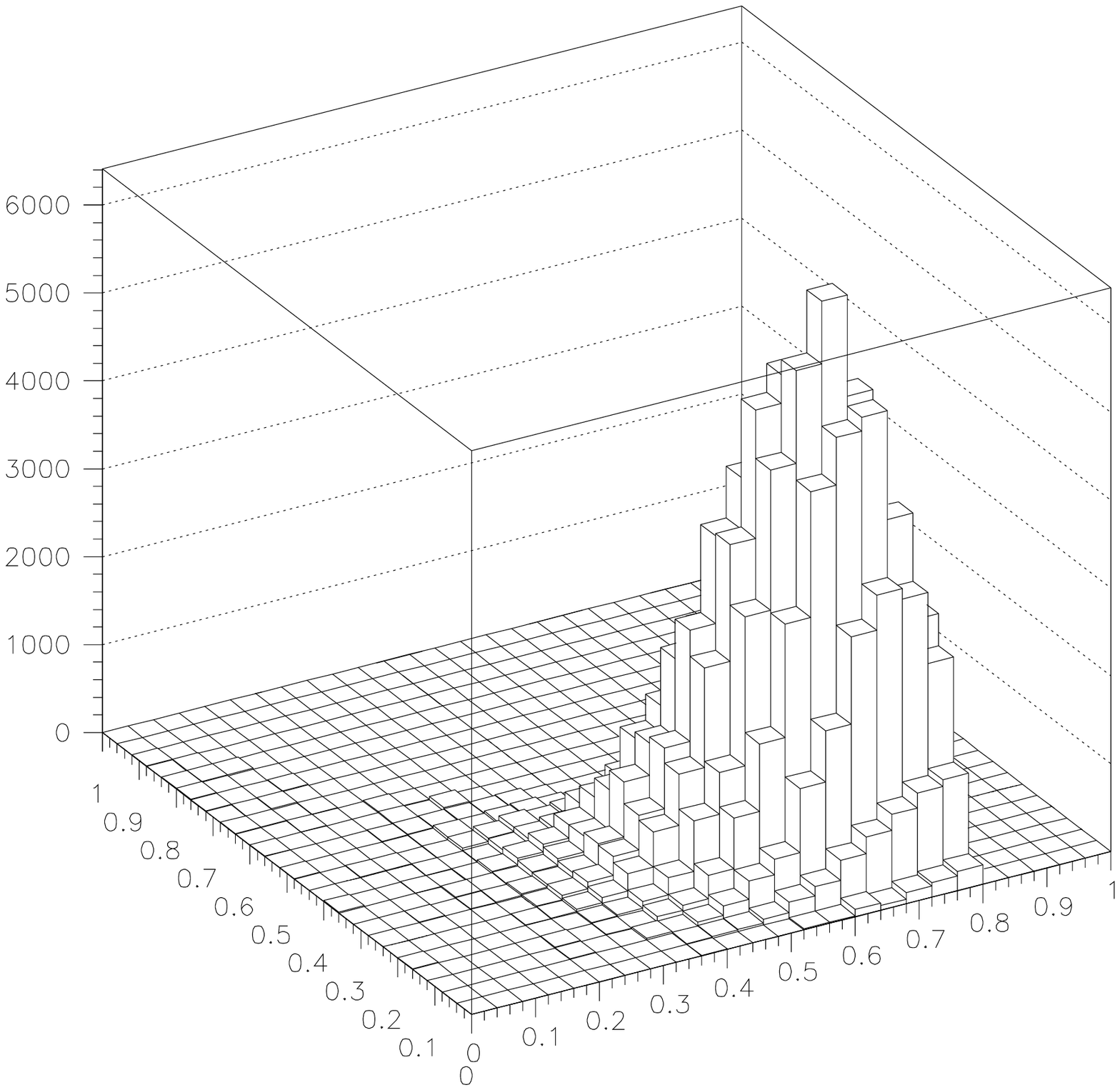}
\vspace{-0.35cm}
\caption{Bidimensional plots in the $X_1$ vs $X_2$ plane.  
The two plots on the left have been obtained for the {\it single-hit} production data,
which contain both the PMT noise and the scintillation events.
The two plots on the right have been obtained for events from a 
$\gamma$ source of suitable strength; thus just scintillation events are present.
The plots on the top refer to the 2-4 keV energy interval and those on the bottom to 4-6 keV.
The noise is clearly separated from the scintillation events; see text.}
\label{fg:lego}
\end{figure}

The noise rejection in the {\it single-hit} events is performed by considering that the 
scintillation pulses in NaI(Tl) have 
a time distribution with decay time of order of hundreds of ns (typically $\simeq$ 240 ns), while the 
noise pulses are single fast photoelectrons pulses with decay time of order of tens of ns (see 
Fig.~\ref{fg:pulses}).
The different time characteristics of these signals can be investigated building several 
variables from the information recorded for each pulse by the Waveform Analyser
over the 2048 ns time window. The variables follow different distributions for 
scintillation and noise pulses.
In particular, the fractions of the pulse areas evaluated over different time 
intervals, such as:
\begin{equation}
X_1 = \frac{ \mbox{Area (from 100 ns to 600 ns)} }{ \mbox{Area (from 0 ns to 600 ns)} }; \, \, \, \,
X_2 = \frac{ \mbox{Area (from 0 ns to 50 ns)}    }{ \mbox{Area (from 0 ns to 600 ns)} },
\end{equation}
offer a very effective procedure\footnote{
Note that the 600 ns time interval corresponds to the usual hardware gate for NaI(Tl) 
pulses and it is equal to $\simeq$ 2.5 times the decay time of a scintillation pulse, that is the area 
(from 0 ns to 600 ns) is proportional to the energy of the event.}.
In fact, the first variable is distributed around 0 for noise events and around 
0.7 for scintillation pulses; the second variable indeed is distributed around 1 for noise and 
around 0.25 for scintillation events\footnote{Qualitatively, hypothesizing only one decay constant 
the expected value for scintillation events should be 0.64 for the first variable and $\simeq$ 0.2 for 
the second one.}. The values of the 
two variables quoted above, obtained for each event of the production data, can be plotted one $vs$ 
the other, as in Fig.~\ref{fg:lego}{\it--left panels};
the signature of the residual noise above the software energy threshold 
with respect to the scintillation events is evident.

Thus, suitable acceptance windows can be applied to obtain just scintillation events. 
Figure ~\ref{fg:lego} clearly points out that stringent acceptance windows can be used
to select a data sample clean from noise;
the effectiveness of this procedure has already been demonstrated by DAMA/NaI \cite{RNC,Sist}.

\vspace{0.8cm} 
\noindent {\bf The overall efficiency for Dark Matter investigation}

As mentioned above, two PMTs on each detector work in
coincidence at single photoelectron threshold. This 
-- considering also the relatively high available 
number of photoelectrons/keV -- assures 
that the required coincidence of the two PMTs does not introduce a very significant 
hardware cut of the events near the energy threshold. In fact, 
it is $\simeq 0.8$ in the 2-3 keV bin and 1 above 5 keV for a coincidence time window of 50 ns
and practically always equal to unity in case of a 100 ns time window.

\begin{figure}[!ht]
\centering
\includegraphics[width=6.cm] {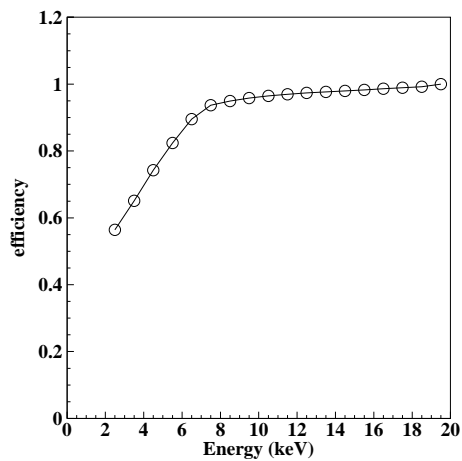}
\caption{Typical behavior of the overall efficiencies for the {\it single-hit} events.
They are dominated by the efficiency related to the stringent acceptance windows used in 
the data analysis to assure the full rejection of noise near energy threshold
(see Fig. \ref{fg:lego}); see text.}
\label{fg:eff}
\end{figure}

The efficiencies related to the used acceptance windows for the {\it single-hit} events in the 
lowest energy bins can be determined and properly used in order to account 
for scintillation events, that might be lost by applying the acceptance windows.
These efficiencies, for each considered energy bin, are measured by applying the same acceptance windows 
to events induced by a $^{241}$Am source of suitable activity in the same experimental conditions 
and energy range as the production data (see Fig.~\ref{fg:lego}{\it--right panels}).
Thus, very stringent acceptance windows, which assure absence of any noise tail, 
can be considered and related efficiencies can be properly evaluated and used
(see Fig. \ref{fg:eff}).
In particular, to determine the efficiencies of the acceptance windows
and their stability over the whole
data taking periods, they are regularly measured by a suitable $^{241}$Am 
source\footnote{The use of $^{137}$Cs Compton electron calibrations 
practically gives the same values as also already verified in DAMA/NaI.}.
These routine efficiency measurements are
performed roughly each 10 days; each
time typically 10$^4$ -- 10$^5$ events per keV are collected.

Let us now consider the radiation which can be induced in the crystal by the Dark Matter particles
(see e.g. \cite{RNC,ijmd,ijma,ijma07,wimpele,LDM}). 
Also thanks to the large number of available photoelectrons and to the single
photoelectron threshold of each PMT working in coincidence, the
detection efficiencies for internal photons and/or X-rays in the few keV range 
approaches the unity as well as for electrons. The detection efficiency for nuclear recoils
is practically equal to unity considering the shortness of the recoil's range in matter.

\vspace{0.1cm} 
\noindent {\bf The energy threshold}

As shown, the new DAMA/LIBRA detectors have been calibrated down to the keV region 
(see Fig. \ref{k40_8} and Fig. \ref{fg:calib}).
This assures a clear knowledge of the ``physical'' energy threshold of the experiment.
It obviously profits of the relatively high number of available photoelectrons/keV; 
the larger this number is, the lower is the reachable physical threshold.
In particular, in DAMA/LIBRA the two PMTs of each detector work in coincidence with hardware threshold
at single photoelectron level, while the software energy threshold used by the experiment is 2 keV.
This latter value is well supported by the keV range calibrations,
by the features of the near-threshold-noise full rejection procedure 
and by the efficiencies when lowering the number of available photoelectrons (see for example 
Fig. \ref{fg:eff}).

\vspace{0.1cm} 
\noindent {\bf The {\it single-hit} scintillation events at low energy}

\begin{figure}[!ht]
\vspace{-0.8cm}
\centering
\includegraphics[width=10.cm] {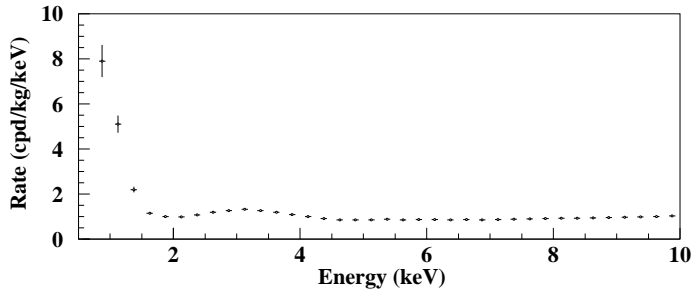}
\vspace{-0.6cm}
\caption{Cumulative low-energy distribution
of the {\it single-hit} scintillation events (that is each detector has all the others as veto)
as measured by the used DAMA/LIBRA 
detectors in an exposure of about 0.53 ton $\times$ yr. 
The energy threshold of the experiment is at 2 keV and corrections for efficiencies are already applied. See text.}
\label{fg:0}
\end{figure}

The procedure for noise rejection near energy threshold, described above, 
is the only procedure applied to the collected data. 
Fig. \ref{fg:0} shows, as example, the resulting cumulative 
low energy distribution of the {\it single-hit} scintillation events, as measured
by DAMA/LIBRA detectors in an exposure of 0.53 ton $\times$ yr.
This energy distribution is the mean value of all the used detectors; some differences among the
detectors are present depending e.g. on their specific levels of residual contaminants 
and on their position in the matrix. 

It is worth noting that -- as in the former DAMA/NaI experiment -- neither other 
on-line or off-line techniques nor backgrounds from residual radioactivity suppression 
are performed at all. 
In particular, background contribution to the counting rate of the {\it single-hit} scintillation 
events, arising from residual radioactivity in the experimental apparatus,
cannot precisely be extrapolated in the keV
region e.g. because of: i) the limitation of 
MonteCarlo simulation 
programs at very low energies; ii) the fact that often just upper limits for residual contaminants are 
available; iii) the unknown location of each residual contaminant
in each component; etc. Nevertheless, the investigations, presented in previous section, 
are extremely useful e.g. to 
qualify the detectors and to identify the sources 
which should be reduced in 
further developments of radiopure crystals, detectors' components, apparatus components, etc.. 
On the other hand, as known, the annual modulation signature, which is exploited by DAMA 
apparata, acts itself as an effective background rejection procedure.

\vspace{0.1cm} 
\noindent {\bf Response to nuclear recoils}

Finally, it is worth noting that, whenever WIMP (or WIMP-like) candidates are considered 
in corollary analyses of Dark Matter investigations, the response of the NaI(Tl) detectors 
to nuclear recoils has to be taken into account \cite{RNC}.
Moreover, for a correct evaluation of the phenomenologies connected with the detection of 
WIMP (or WIMP-like) candidates, the so-called Migdal effect \cite{ijma07} and 
the channeling effects \cite{chan} have also to be considered.
Detailed discussions can be found e.g. in ref. \cite{ijma07,chan}.

\vspace{0.1cm} 
\noindent {\bf Stability of the running conditions}

As already done for DAMA/NaI (see e.g. ref. \cite{Sist,RNC}),
the stability of several parameters and of the running conditions 
is monitored along all the data taking. This aspect is directly related
with the experimental results on the annual modulation signature and,
therefore, it will be discussed elsewhere.

\section{Conclusions}

In this paper the $\simeq$ 250 kg highly radiopure NaI(Tl) DAMA/LIBRA 
apparatus, running at the Gran Sasso National Laboratory of the I.N.F.N., has been described.
The main components and performances have been discussed.

\end{document}